\documentclass[12pt]{article}
\usepackage{amssymb}
\begin{document}



\def\a{\alpha}
\def\b{\beta}
\def\d{\delta}
\def\e{\epsilon}
\def\g{\gamma}
\def\h{\mathfrak{h}}
\def\k{\kappa}
\def\l{\lambda}
\def\o{\omega}
\def\p{\wp}
\def\r{\rho}
\def\t{\tau}
\def\s{\sigma}
\def\z{\zeta}
\def\x{\xi}
 \def\A{{\cal{A}}}
 \def\B{{\cal{B}}}
 \def\C{{\cal{C}}}
 \def\D{{\cal{D}}}
\def\G{\Gamma}
\def\K{{\cal{K}}}
\def\O{\Omega}
\def\R{\bar{R}}
\def\T{{\cal{T}}}
\def\L{\Lambda}
\def\f{E_{\tau,\eta}(sl_2)}
\def\E{E_{\tau,\eta}(sl_n)}
\def\Zb{\mathbb{Z}}
\def\Cb{\mathbb{C}}

\def\R{\overline{R}}

\def\beq{\begin{equation}}
\def\eeq{\end{equation}}
\def\bea{\begin{eqnarray}}
\def\eea{\end{eqnarray}}
\def\ba{\begin{array}}
\def\ea{\end{array}}
\def\no{\nonumber}
\def\le{\langle}
\def\re{\rangle}
\def\lt{\left}
\def\rt{\right}

\newtheorem{Theorem}{Theorem}
\newtheorem{Definition}{Definition}
\newtheorem{Proposition}{Proposition}
\newtheorem{Lemma}{Lemma}
\newtheorem{Corollary}{Corollary}
\newcommand{\proof}[1]{{\bf Proof. }
        #1\begin{flushright}$\Box$\end{flushright}}

\baselineskip=20pt

\newfont{\elevenmib}{cmmib10 scaled\magstep1}
\newcommand{\preprint}{
   \begin{flushleft}
   \end{flushleft}\vspace{-1.3cm}
   \begin{flushright}\normalsize
   \end{flushright}}
\newcommand{\Title}[1]{{\baselineskip=26pt
   \begin{center} \Large \bf #1 \\ \ \\ \end{center}}}
\newcommand{\Author}{\begin{center}
   \large \bf
Wen-Li Yang${}^{a}$, ~Xi Chen${}^{a}$, ~Jun Feng${}^{a}$,~Kun Hao${}^{a}$,~Ke Wu ${}^b$,
~Zhan-Ying Yang${}^c$ ~and~Yao-Zhong Zhang ${}^d$
 \end{center}}
\newcommand{\Address}{\begin{center}

     ${}^a$ Institute of Modern Physics, Northwest University,
     Xian 710069, P.R. China\\
     ${}^b$ School of Mathematical Science, Capital Normal University,
     Beijing 100037, P.R. China \\
     ${}^c$ The Department of Physics, Northwest University,
     Xian 710069, P.R. China \\
     ${}^d$ The University of Queensland, School of Mathematics and Physics,  Brisbane, QLD 4072,
     Australia\\
   \end{center}}
\preprint
\thispagestyle{empty}
\bigskip\bigskip\bigskip

\Title{Scalar products of the open
XYZ chain with non-diagonal boundary terms } \Author

\Address
\vspace{1cm}

\begin{abstract}
With the help of the F-basis provided by the Drinfeld twist or factorizing
F-matrix of the  eight-vertex solid-on-solid (SOS) model, we obtain the determinant representations of the scalar
products of Bethe states for the open XYZ chain  with non-diagonal boundary terms. By taking the on shell limit,
we obtain the determinant representations (or Gaudin formula)
of the norms of the Bethe states.

\vspace{1truecm} \noindent {\it PACS:}  75.10.Pq; 04.20.Jb;
05.50.+q

\noindent {\it Keywords}: Spin chain; Bethe
ansatz; Scalar products.
\end{abstract}
\newpage
\section{Introduction}
\label{intro} \setcounter{equation}{0}

One of the most challenging and important problems in theory of
quantum integrable models, having obtained the spectrum and
eigenstates of the corresponding Hamiltonians, is to construct
exact and manageable expressions of correlation functions (or scalar products of Bethe states)
\cite{Smi92, Kor93}. This
problem is also fundamental to enlarge the range of applications
of these models in the realm of condensed matter physics, quantum
information theory. There are two approaches in the literature for computing the correlation
functions of a quantum integrable model. One is the vertex operator method (see e.g.
\cite{Fre92,Dav93,Koy94,Hou97,Yan99,Hou99}) which works only on an infinite lattice,
and another one is based on the detailed analysis of the structure of the Bethe states
\cite{Kor82, Ize87}. As for the second approach which usually works for models with
finite size, it is well known that
in the framework of quantum inverse scattering method (QISM) \cite{Kor93} Bethe states are
obtained by applying pseudo-particle creation operators to
reference state (pseudo-vacuum). However, the apparently simple action of creation operators is
plagued with non-local effects arising from polarization clouds or
compensating exchange terms on the level of local operators. This makes the direct
calculation of correlation functions of models with finite size challenging.

Progress has recently
been made on the second approach with the help of the Drinfeld twists or
factorizing F-matrices \cite{Dri83}. Working in the F-basis provided by the F-matrices,
the authors in \cite{Mai00, Kit99} managed to calculate the form factors and correlation
functions of the XXX and XXZ chains  with periodic boundary condition (or closed chains)
analytically and expressed them in determinant forms. Then the determinant representation
of the scalar products and correlation functions of the supersymmetric t-J model \cite{Zha06}
and its q-deformed model \cite{Zha06-1} with periodic boundary condition was obtained within the
corresponding F-basis given in \cite{Yan04-2}.

It was  noticed  \cite{Wan02,Kit07} that the F-matrices of
the closed XXX and XXZ chains also  make the pseudo-particle
creation operators of the open XXX and XXZ chains with diagonal
boundary terms polarization free. This is mainly due to the fact
that the closed chain and the corresponding open chain with
diagonal boundary terms share the same reference state
\cite{Skl88}. However, the story for the open XXZ chain with
non-diagonal boundary terms is quite different
\cite{Nep04,Cao03,Yan04,Gal05,Gie05,Gie05-1,Yan04-1,Baj06,Yan05,Doi06,Mur06,Bas07,Gal08,Mur09,Ami10,Cra10}.
Firstly, the reference state (all spin up state) of the closed
chain is no longer a reference state of the open chain with
non-diagonal boundary terms \cite{Cao03,Yan04,Yan04-1}. Secondly,
at least two reference states (and thus two sets of Bethe states) are
needed \cite{Yan07} for the open XXZ chain with non-diagonal
boundary terms in order to obtain its complete spectrum
\cite{Nep03,Yan06}. As a consequence, the F-matrix found in
\cite{Mai00} is no longer the {\it desirable} F-matrix for the
open XXZ chain with non-diagonal boundary terms. With the help of
the F-matrices of SOS models given in \cite{Alb00,Yan10}, the domain wall (DW) partition
function of the six-vertex model with a non-diagonal reflecting end \cite{Tsu98,Yan10-1}
(or the DW partition function of the trigonometric
SOS model with reflecting end \cite{Fil10}), the explicit and completely symmetric
expressions \cite{Yan10} of the two sets of Bethe states and the
determinant representation \cite{Yan11-1} of scalar products of these Bethe states   for the open XXZ chain with non-diagonal
boundary terms have been obtained.

It is well known that among solvable models elliptic ones stand out as a particularly
important class due to the fact that most trigonometric and
rational models can be obtained from them by  certain limits. In
this paper, we focus on the most fundamental elliptic model---the
XYZ spin chain  \cite{Bax82} whose trigonometric/rational limit
gives the XXZ/XXX chain. Basing the work \cite{Alb00} we
have succeeded in obtaining the DW partition
function of the eight-vertex model with a non-diagonal reflecting end \cite{Yan11} (or
the DW partition function of the eight-vertex
SOS model with reflecting end \cite{Fil10-1}) and
the explicit and completely symmetric
expressions of the two sets of Bethe states of the open XYZ chain with
non-diagonal boundary terms \cite{Yan11-2}.  In this paper,
we shall investigate  the determinant representations of scalar products
of the Bethe states of the open XYZ chain with non-diagonal boundary terms specified by the non-diagonal
K-matrices (\ref{K-matrix-2-1}) and (\ref{K-matrix-6}) given by \cite{Ina94,Hou95}.

The paper is organized as follows.  In section 2, we briefly
describe the open XYZ chain with non-diagonal boundary terms, and
introduce the pseudo-particle creation operators and the two sets
of Bethe states of the model. In section 3, we introduce the face picture
of the model and express the scalar products in terms of the operators in the
face picture. In section 4, we use the F-matrix of the eight-vertex SOS model
to construct   the completely symmetric and polarization free representations of the
pseudo-particle creation/annihilation operators  in the F-basis. In the face picture,
with the help of the F-basis, we obtain the determinant representations of the scalar
products of Bethe states and the determinant representations (or Gaudin formula)
of the norms of the Bethe states in section 5. In
section 6, we summarize our results and give some discussions.


\section{ The inhomogeneous spin-$\frac{1}{2}$ XYZ open chain}
\label{XYZ} \setcounter{equation}{0}

Let us fix $\tau$ such that ${\rm Im}(\tau)>0$ and a generic complex number $\eta$.
Introduce the following elliptic functions
\bea
\theta\lt[\begin{array}{c} a\\b
  \end{array}\rt](u,\tau)&=&\sum_{n=-\infty}^{\infty}
  \exp\lt\{i\pi\lt[(n+a)^2\tau+2(n+a)(u+b)\rt]\rt\},\label{Function-a-b}\\
\theta^{(j)}(u)&=&\theta\lt[\begin{array}{c}\frac{1}{2}-\frac{j}{2}\\
 [2pt]\frac{1}{2}
 \end{array}\rt](u,2\tau),\quad j=1,2;\qquad
 \s(u)=\theta\lt[\begin{array}{c}\frac{1}{2}\\[2pt]\frac{1}{2}
 \end{array}\rt](u,\tau).
 \label{Function-j}\eea The
$\s$-function\footnote{Our $\s$-function is the
$\vartheta$-function $\vartheta_1(u)$ \cite{Whi50}. It has the
following relation with the {\it Weierstrassian\/} $\s$-function
$\s_w(u)$: $\s_w(u)\propto e^{\eta_1u^2}\s(u)$ with
$\eta_1=\pi^2(\frac{1}{6}-4\sum_{n=1}^{\infty}\frac{nq^{2n}}{1-q^{2n}})
$ and $q=e^{i\tau}$.}
 satisfies the so-called Riemann
identity:\bea
&&\s(u+x)\s(u-x)\s(v+y)\s(v-y)-\s(u+y)\s(u-y)\s(v+x)\s(v-x)\no\\
&&~~~~~~=\s(u+v)\s(u-v)\s(x+y)\s(x-y),\label{identity}\eea which
will be useful in the following. Moreover, for any $\a=(\a_1,\a_2),\,\a_1,\a_2\in\Zb_2$, we
can introduce a function $\s_{\a}(u)$ as follow
\bea
  \s_{\a}(u)=\theta\lt[\begin{array}{c}\frac{1}{2}+\frac{\a_1}{2}\\[2pt]\frac{1}{2}+\frac{\a_2}{2}
 \end{array}\rt](u,\tau),\quad\quad \a_1,\a_2\in \Zb_2.\label{sigma-function}
\eea The above definition implies the identification
  $\s_{(0,0)}(u)=\s(u)$.

Let $V$ be a two-dimensional vector space $\Cb^2$ and
$\{\e_i|i=1,2\}$ be the orthonormal basis  of $V$ such that
$\langle \e_i,\e_j\rangle=\d_{ij}$. The well-known eight-vertex
model R-matrix $\R(u)\in {\rm End}(V\otimes V)$ is given by \bea
\R(u)=\lt(\begin{array}{llll}a(u)&&&d(u)\\&b(u)&c(u)&\\
&c(u)&b(u)&\\d(u)&&&a(u)\end{array}\rt). \label{r-matrix}\eea The
non-vanishing matrix elements  are \cite{Bax82}\bea
&&a(u)=\frac{\theta^{(1)}(u)\,\theta^{(0)}(u+\eta)\,\s(\eta)}
{\theta^{(1)}(0)\, \theta^{(0)}(\eta)\,\s(u+\eta)},\quad
b(u)=\frac{\theta^{(0)}(u)\,
 \theta^{(1)}(u+\eta)\,\s(\eta)}
{\theta^{(1)}(0)\,\theta^{(0)}(\eta)\,\s(u+\eta)},\no\\[6pt]
&&c(u)=\frac{\theta^{(1)}(u)\,
 \theta^{(1)}(u+\eta)\,\s(\eta)}
{\theta^{(1)}(0)\, \theta^{(1)}(\eta)\,\s(u+\eta)},\quad
d(u)=\frac{\theta^{(0)}(u)\,
 \theta^{(0)}(u+\eta)\,\s(\eta)}
{\theta^{(1)}(0)\theta^{(1)}(\eta)\,\s(u+\eta)}.\label{r-func}\eea Here
$u$ is the spectral parameter and $\eta$ is the so-called crossing
parameter. The R-matrix satisfies the quantum Yang-Baxter equation
(QYBE)
\bea \R_{1,2}(u_1-u_2)\R_{1,3}(u_1-u_3)\R_{2,3}(u_2-u_3)
=\R_{2,3}(u_2-u_3)\R_{1,3}(u_1-u_3)\R_{1,2}(u_1-u_2).\label{QYBE}\eea
Throughout we adopt the standard notation: for any
matrix $A\in {\rm End}(V)$, $A_j$ (or  $A^j$)is an embedding operator in the
tensor space $V\otimes V\otimes\cdots$, which acts as $A$ on the
$j$-th space and as identity on the other factor spaces;
$R_{i,j}(u)$ is an embedding operator of R-matrix in the tensor
space, which acts as identity on the factor spaces except for the
$i$-th and $j$-th ones.

One introduces the ``row-to-row"  (or one-row ) monodromy matrix
$T(u)$, which is an $2\times 2$ matrix with elements being
operators acting on $V^{\otimes N}$, where $N=2M$ ($M$ being a
positive integer),\bea
T_0(u)=\R_{0,N}(u-z_N)\R_{0,N-1}(u-z_{N-1})\cdots
\R_{0,1}(u-z_1).\label{Mon-V}\eea Here $\{z_j|j=1,\cdots,N\}$ are
arbitrary free complex parameters which are usually called
inhomogeneous parameters.

Integrable open chain can be constructed as follows \cite{Skl88}.
Let us introduce a pair of K-matrices $K^-(u)$ and $K^+(u)$. The
former satisfies the reflection equation (RE)
 \bea &&\R_{1,2}(u_1-u_2)K^-_1(u_1)\R_{2,1}(u_1+u_2)K^-_2(u_2)\no\\
 &&~~~~~~=
K^-_2(u_2)\R_{1,2}(u_1+u_2)K^-_1(u_1)\R_{2,1}(u_1-u_2),\label{RE-V}\eea
and the latter  satisfies the dual RE \bea
&&\R_{1,2}(u_2-u_1)K^+_1(u_1)\R_{2,1}(-u_1-u_2-2\eta)K^+_2(u_2)\no\\
&&~~~~~~=
K^+_2(u_2)\R_{1,2}(-u_1-u_2-2\eta)K^+_1(u_1)\R_{2,1}(u_2-u_1).
\label{DRE-V}\eea For open spin-chains, instead of the standard
``row-to-row" monodromy matrix $T(u)$ (\ref{Mon-V}), one needs to
consider  the
 ``double-row" monodromy matrix $\mathbb{T}(u)$
\bea
  \mathbb{T}(u)=T(u)K^-(u)\hat{T}(u),\quad \hat{T}(u)=T^{-1}(-u).
  \label{Mon-V-0}
\eea Then the double-row transfer matrix of the XYZ chain with
open boundary (or the open XYZ chain) is given by
\bea
  \t(u)=tr(K^+(u)\mathbb{T}(u)).\label{trans}
\eea The QYBE and
(dual) REs lead to that the transfer matrices with different
spectral parameters commute with each other \cite{Skl88}:
$[\t(u),\t(v)]=0$. This ensures the integrability of the open XYZ
chain.

In this paper, we consider the K-matrix $K^{-}(u)$ which is a
generic solution \cite{Ina94,Hou95} to the RE (\ref{RE-V}) associated with the  R-matrix
(\ref{r-matrix})
\bea K^-(u)=k^-_0(u)+k^-_x(u)\,\s^x+k^-_y\,\s^y+k^-_z(u)\,\s^z,\label{K-matrix}\eea
where $\s^x,\s^y,\s^z$ are the Pauli matrices and the coefficient functions are
\bea
&& k^-_0(u)=\frac{\s(2u)\,\s(\l_1+\l_2-\frac{1}{2})\,\s(\l_1+\xi)\,\s(\l_2+\xi)}
   {2\,\s(u)\,\s(-u+\l_1+\l_2-\frac{1}{2})\,\s(\l_1+\xi+u)\,\s(\l_2+\xi+u)},\no\\[6pt]
&& k^-_x(u)=\frac{\s(2u)\,\s_{(1,0)}(\l_1+\l_2-\frac{1}{2})\,\s_{(1,0)}(\l_1+\xi)\,\s_{(1,0)}(\l_2+\xi)}
   {2\,\s_{(1,0)}(u)\,\s(-u+\l_1+\l_2-\frac{1}{2})\,\s(\l_1+\xi+u)\,\s(\l_2+\xi+u)},\no\\[6pt]
&& k^-_y(u)=\frac{i\,\s(2u)\,\s_{(1,1)}(\l_1+\l_2-\frac{1}{2})\,\s_{(1,1)}(\l_1+\xi)\,\s_{(1,1)}(\l_2+\xi)}
   {2\,\s_{(1,1)}(u)\,\s(-u+\l_1+\l_2-\frac{1}{2})\,\s(\l_1+\xi+u)\,\s(\l_2+\xi+u)},\no\\[6pt]
&& k^-_z(u)=\frac{\s(2u)\,\s_{(0,1)}(\l_1+\l_2-\frac{1}{2})\,\s_{(0,1)}(\l_1+\xi)\,\s_{(0,1)}(\l_2+\xi)}
   {2\,\s_{(0,1)}(u)\,\s(-u+\l_1+\l_2-\frac{1}{2})\,\s(\l_1+\xi+u)\,\s(\l_2+\xi+u)}.\label{K-matrix-2-1} \eea
At the same time, we introduce  the corresponding {\it dual\/}
K-matrix $K^+(u)$ which is a generic solution to the dual
reflection equation (\ref{DRE-V}) with a particular choice of the
free boundary parameters according to those of $K^-(u)$ (\ref{K-matrix})-(\ref{K-matrix-2-1}):
\bea
 K^+(u)=k^+_0(u)+k^+_x(u)\,\s^x+k^+_y\,\s^y+k^+_z(u)\,\s^z,
 \label{DK-matrix}
\eea with the coefficient functions
\bea
&& k^+_0(u)=\frac{\s(\hspace{-0.06truecm}-\hspace{-0.06truecm}2u\hspace{-0.06truecm}-\hspace{-0.06truecm}2\eta)
   \s(\l_1\hspace{-0.06truecm}+\hspace{-0.06truecm}\l_2\hspace{-0.06truecm}+\hspace{-0.06truecm}\eta
   \hspace{-0.06truecm}-\hspace{-0.06truecm}\frac{1}{2})
   \s(\l_1\hspace{-0.06truecm}+\hspace{-0.06truecm}\bar{\xi})
   \s(\l_2\hspace{-0.06truecm}+\hspace{-0.06truecm}\bar{\xi})}
   {2\s(\hspace{-0.06truecm}-\hspace{-0.06truecm}u\hspace{-0.06truecm}-\hspace{-0.06truecm}\eta)
   \s(u\hspace{-0.06truecm}+\hspace{-0.06truecm}\eta\hspace{-0.06truecm}+\hspace{-0.06truecm}\l_1\hspace{-0.06truecm}+\hspace{-0.06truecm}\l_2
   \hspace{-0.06truecm}-\hspace{-0.06truecm}\frac{1}{2})\s(\l_1\hspace{-0.06truecm}+\hspace{-0.06truecm}\bar{\xi}
   \hspace{-0.06truecm}-\hspace{-0.06truecm}u\hspace{-0.06truecm}-\hspace{-0.06truecm}\eta)
   \s(\l_2\hspace{-0.06truecm}+\hspace{-0.06truecm}\bar{\xi}\hspace{-0.06truecm}-\hspace{-0.06truecm}u\hspace{-0.06truecm}-\hspace{-0.06truecm}\eta)},\no\\[6pt]
&& k^+_x(u)=\frac{\s(\hspace{-0.06truecm}-\hspace{-0.06truecm}2u\hspace{-0.06truecm}-\hspace{-0.06truecm}2\eta)
   \s_{(1,0)}(\l_1\hspace{-0.06truecm}+\hspace{-0.06truecm}\l_2\hspace{-0.06truecm}+\hspace{-0.06truecm}\eta
   \hspace{-0.06truecm}-\hspace{-0.06truecm}\frac{1}{2})\s_{(1,0)}(\l_1\hspace{-0.06truecm}+\hspace{-0.06truecm}\bar{\xi})
   \s_{(1,0)}(\l_2\hspace{-0.06truecm}+\hspace{-0.06truecm}\bar{\xi})}
   {2\s_{(1,0)}(\hspace{-0.06truecm}-\hspace{-0.06truecm}u\hspace{-0.06truecm}-\hspace{-0.06truecm}\eta)
   \s(u\hspace{-0.06truecm}+\hspace{-0.06truecm}\eta\hspace{-0.06truecm}+\hspace{-0.06truecm}\l_1\hspace{-0.06truecm}+\hspace{-0.06truecm}\l_2
   \hspace{-0.06truecm}-\hspace{-0.06truecm}\frac{1}{2})\s(\l_1\hspace{-0.06truecm}+\hspace{-0.06truecm}\bar{\xi}
   \hspace{-0.06truecm}-\hspace{-0.06truecm}u\hspace{-0.06truecm}-\hspace{-0.06truecm}\eta)
   \s(\l_2\hspace{-0.06truecm}+\hspace{-0.06truecm}\bar{\xi}\hspace{-0.06truecm}-\hspace{-0.06truecm}u\hspace{-0.06truecm}-\hspace{-0.06truecm}\eta)},\no\\[6pt]
&& k^+_y(u)=\frac{i\,\s(-\hspace{-0.06truecm}2u\hspace{-0.06truecm}-\hspace{-0.06truecm}2\eta)
   \s_{(1,1)}(\l_1\hspace{-0.06truecm}+\hspace{-0.06truecm}\l_2\hspace{-0.06truecm}+\hspace{-0.06truecm}\eta\hspace{-0.06truecm}-\hspace{-0.06truecm}\frac{1}{2})
   \s_{(1,1)}(\l_1\hspace{-0.06truecm}+\hspace{-0.06truecm}\bar{\xi})\s_{(1,1)}(\l_2\hspace{-0.06truecm}+\hspace{-0.06truecm}\bar{\xi})}
   {2\s_{(1,1)}(\hspace{-0.06truecm}-\hspace{-0.06truecm}u\hspace{-0.06truecm}-\hspace{-0.06truecm}\eta)
   \s(u\hspace{-0.06truecm}+\hspace{-0.06truecm}\eta\hspace{-0.06truecm}+\hspace{-0.06truecm}\l_1\hspace{-0.06truecm}+\hspace{-0.06truecm}
   \l_2\hspace{-0.06truecm}-\hspace{-0.06truecm}\frac{1}{2})
   \s(\l_1\hspace{-0.06truecm}+\hspace{-0.06truecm}\bar{\xi}\hspace{-0.06truecm}-\hspace{-0.06truecm}u\hspace{-0.06truecm}-\hspace{-0.06truecm}\eta)
   \s(\l_2\hspace{-0.06truecm}+\hspace{-0.06truecm}\bar{\xi}\hspace{-0.06truecm}-\hspace{-0.06truecm}u\hspace{-0.06truecm}-\hspace{-0.06truecm}\eta)},\no\\[6pt]
&& k^+_z(u)=\frac{\s(\hspace{-0.06truecm}-\hspace{-0.06truecm}2u\hspace{-0.06truecm}-\hspace{-0.06truecm}2\eta)
   \s_{(0,1)}(\l_1\hspace{-0.06truecm}+\hspace{-0.06truecm}\l_2\hspace{-0.06truecm}+\hspace{-0.06truecm}\eta\hspace{-0.06truecm}-\hspace{-0.06truecm}\frac{1}{2})
   \s_{(0,1)}(\l_1\hspace{-0.06truecm}+\hspace{-0.06truecm}\bar{\xi})\s_{(0,1)}(\l_2\hspace{-0.06truecm}+\hspace{-0.06truecm}\bar{\xi})}
   {2\s_{(0,1)}(\hspace{-0.06truecm}-\hspace{-0.06truecm}u\hspace{-0.06truecm}-\hspace{-0.06truecm}\eta)
   \s(u\hspace{-0.06truecm}+\hspace{-0.06truecm}\eta\hspace{-0.06truecm}+\hspace{-0.06truecm}\l_1\hspace{-0.06truecm}+\hspace{-0.06truecm}\l_2
   \hspace{-0.06truecm}-\hspace{-0.06truecm}\frac{1}{2})
   \s(\l_1\hspace{-0.06truecm}+\hspace{-0.06truecm}\bar{\xi}\hspace{-0.06truecm}-\hspace{-0.06truecm}u\hspace{-0.06truecm}-\hspace{-0.06truecm}\eta)
   \s(\l_2\hspace{-0.06truecm}+\hspace{-0.06truecm}\bar{\xi}\hspace{-0.06truecm}-\hspace{-0.06truecm}u\hspace{-0.06truecm}-\hspace{-0.06truecm}\eta)}.\label{K-matrix-6}
\eea The K-matrices $K^{\mp}(u)$ depend on four free boundary parameters
$\{\l_1,\,\l_2,\,\xi,\,\bar{\xi}\}$. It is
very convenient to introduce a vector $\l\in V$ associated with
the boundary parameters $\{\l_i\}$, \bea
 \l=\sum_{k=1}^2\l_k\e_k. \label{boundary-vector}
\eea


\subsection{Vertex-face correspondence}

Let us briefly review the face-type R-matrix associated with the
six-vertex model.

Set \bea \hat{\imath}=\e_i-\overline{\e},~~\overline{\e}=
\frac{1}{2}\sum_{k=1}^{2}\e_k, \quad i=1,2,\qquad {\rm then}\,
\sum_{i=1}^2\hat{\imath}=0. \label{fundmental-vector} \eea Let
$\h$ be the Cartan subalgebra of $A_{1}$ and $\h^{*}$ be its dual.
A finite dimensional diagonalizable  $\h$-module is a complex
finite dimensional vector space $W$ with a weight decomposition
$W=\oplus_{\mu\in \h^*}W[\mu]$, so that $\h$ acts on $W[\mu]$ by
$x\,v=\mu(x)\,v$, $(x\in \h,\,v\in\,W[\mu])$. For example, the
non-zero weight spaces of the fundamental representation
$V_{\L_1}=\Cb^2=V$ are
\bea
 W[\hat{\imath}]=\Cb \e_i,~i=1,2.\label{Weight}
\eea

For a generic $m\in V$, define \bea m_i=\langle m,\e_i\rangle,
~~m_{ij}=m_i-m_j=\langle m,\e_i-\e_j\rangle,~~i,j=1,2.
\label{Def1}\eea Let $R(u,m)\in {\rm End}(V\otimes V)$ be the
R-matrix of the eight-vertex SOS model \cite{Bax82} given by
\bea
R(u;m)\hspace{-0.1cm}=\hspace{-0.1cm}
\sum_{i=1}^{2}R(u;m)^{ii}_{ii}E_{ii}\hspace{-0.1cm}\otimes\hspace{-0.1cm}
E_{ii}\hspace{-0.1cm}+\hspace{-0.1cm}\sum_{i\ne
j}^2\lt\{R(u;m)^{ij}_{ij}E_{ii}\hspace{-0.1cm}\otimes\hspace{-0.1cm}
E_{jj}\hspace{-0.1cm}+\hspace{-0.1cm}
R(u;m)^{ji}_{ij}E_{ji}\hspace{-0.1cm}\otimes\hspace{-0.1cm}
E_{ij}\rt\}, \label{R-matrix} \eea where $E_{ij}$ is the matrix
with elements $(E_{ij})^l_k=\d_{jk}\d_{il}$. The coefficient
functions are \bea
 &&R(u;m)^{ii}_{ii}=1,~~
   R(u;m)^{ij}_{ij}=\frac{\s(u)\s(m_{ij}-\eta)}
   {\s(u+\eta)\s(m_{ij})},~~i\neq j,\label{Elements1}\\
 && R(u;m)^{ji}_{ij}=\frac{\s(\eta)\s(u+m_{ij})}
    {\s(u+\eta)\s(m_{ij})},~~i\neq j,\label{Elements2}
\eea  and $m_{ij}$ is defined in (\ref{Def1}). The R-matrix
satisfies the dynamical (modified) quantum Yang-Baxter equation
(or the star-triangle relation) \cite{Bax82}
\begin{eqnarray}
&&R_{1,2}(u_1-u_2;m-\eta h^{(3)})R_{1,3}(u_1-u_3;m)
R_{2,3}(u_2-u_3;m-\eta h^{(1)})\no\\
&&\qquad =R_{2,3}(u_2-u_3;m)R_{1,3}(u_1-u_3;m-\eta
h^{(2)})R_{1,2}(u_1-u_2;m).\label{MYBE}
\end{eqnarray}
Here we have adopted
\bea R_{1,2}(u,m-\eta h^{(3)})\,v_1\otimes
v_2 \otimes v_3=\lt(R(u,m-\eta\mu)\otimes {\rm id }\rt)v_1\otimes
v_2 \otimes v_3,\quad {\rm if}\, v_3\in W[\mu]. \label{Action}
\eea Moreover, one may check that the R-matrix satisfies  the weight
conservation condition, \bea
  \lt[h^{(1)}+h^{(2)},\,R_{1,2}(u;m)\rt]=0,\label{Conservation}
\eea the unitary condition, \bea
 R_{1,2}(u;m)\,R_{2,1}(-u;m)={\rm id}\otimes {\rm
 id},\label{Unitary}
\eea and the crossing relation \bea
 R(u;m)^{kl}_{ij}=\varepsilon_{l}\,\varepsilon_{j}
   \frac{\s(u)\s((m-\eta\hat{\imath})_{21})}
   {\s(u+\eta)\s(m_{21})}R(-u-\eta;m-\eta\hat{\imath})
   ^{\bar{j}\,k}_{\bar{l}\,i},\label{Crossing}
\eea where
\bea \varepsilon_{1}=1,\,\varepsilon_{2}=-1,\quad {\rm
and}\,\, \bar{1}=2,\,\bar{2}=1.\label{Parity} \eea

Let us introduce two intertwiners which are
$2$-component  column vectors $\phi_{m,m-\eta\hat{\jmath}}(u)$
labelled by $\hat{1},\,\hat{2}$. The $k$-th element of
$\phi_{m,m-\eta\hat{\jmath}}(u)$ is given by \bea
\phi^{(k)}_{m,m-\eta\hat{\jmath}}(u)=\theta^{(k)}(u+2m_j),\label{Intvect}\eea
where the functions $\theta^{(j)}(u)$ are given in (\ref{Function-j}).
Explicitly,
\bea \phi_{m,m-\eta\hat{1}}(u)=
\lt(\begin{array}{c}\theta^{(1)}(u+2m_1)\\[6pt]\theta^{(2)}(u+2m_1)\end{array}\rt),\qquad
\phi_{m,m-\eta\hat{2}}(u)=
\lt(\begin{array}{c}\theta^{(1)}(u+2m_2)\\[6pt]\theta^{(2)}(u+2m_2)\end{array}\rt).\eea
One can prove the following identity \cite{Fan98}
\bea
 {\rm det}\lt|\begin{array}{cc}\theta^{(1)}(u+2m_1)&\theta^{(1)}(u+2m_2)\\[6pt]
   \theta^{(2)}(u+2m_1)&\theta^{(2)}(u+2m_2)\end{array}\rt|=C(\tau)\,\s(u+m_1+m_2-\frac{1}{2})
   \,\s(m_{12}),\no
\eea where $C(\tau)$ is non-vanishing constant which only depends on  $\tau$. This implies that
the two intertwiner vectors
$\phi_{m,m-\eta\hat{\imath}}(u)$ are linearly {\it independent}
for a generic $m\in V$.

 Using the intertwiner vectors, one can derive the following face-vertex
correspondence relation \cite{Bax82}\bea &&\R_{1,2}(u_1-u_2)
\phi^1_{m,m-\eta\hat{\imath}}(u_1)
\phi^2_{m-\eta\hat{\imath},m-\eta(\hat{\imath}+\hat{\jmath})}(u_2)
\no\\&&~~~~~~= \sum_{k,l}R(u_1-u_2;m)^{kl}_{ij}
\phi^1_{m-\eta\hat{l},m-\eta(\hat{l}+\hat{k})}(u_1)
\phi^2_{m,m-\eta\hat{l}}(u_2). \label{Face-vertex} \eea  Then the
QYBE (\ref{QYBE}) of the vertex-type R-matrix $\R(u)$ is equivalent
to the dynamical Yang-Baxter equation (\ref{MYBE}) of the SOS
R-matrix $R(u,m)$. For a generic $m$, we can introduce other types
of intertwiners $\bar{\phi},~\tilde{\phi}$ which  are both row
vectors and satisfy the following conditions, \bea
  &&\bar{\phi}_{m,m-\eta\hat{\mu}}(u)
     \,\phi_{m,m-\eta\hat{\nu}}(u)=\d_{\mu\nu},\quad
     \tilde{\phi}_{m+\eta\hat{\mu},m}(u)
     \,\phi_{m+\eta\hat{\nu},m}(u)=\d_{\mu\nu},\label{Int2}\eea
{}from which one  can derive the relations,
\begin{eqnarray}
&&\sum_{\mu=1}^2\phi_{m,m-\eta\hat{\mu}}(u)\,
 \bar{\phi}_{m,m-\eta\hat{\mu}}(u)={\rm id},\label{Int3}\\
&&\sum_{\mu=1}^2\phi_{m+\eta\hat{\mu},m}(u)\,
 \tilde{\phi}_{m+\eta\hat{\mu},m}(u)={\rm id}.\label{Int4}
\end{eqnarray}
With the help of (\ref{Face-vertex})-(\ref{Int4}), we obtain,
\begin{eqnarray}
 &&\tilde{\phi}^1_{m+\eta\hat{k},m}(u_1)\,\R_{1,2}(u_1-u_2)
 \phi^2_{m+\eta\hat{\jmath},m}(u_2)\no\\
 &&\qquad\quad= \sum_{i,l}R(u_1-u_2;m)^{kl}_{ij}\,
 \tilde{\phi}^1_{m+\eta(\hat{\imath}+\hat{\jmath}),m+\eta\hat{\jmath}}(u_1)
 \phi^2_{m+\eta(\hat{k}+\hat{l}),m+\eta\hat{k}}(u_2),\label{Face-vertex1}\\
 &&\tilde{\phi}^1_{m+\eta\hat{k},m}(u_1)
 \tilde{\phi}^2_{m+\eta(\hat{k}+\hat{l}),m+\eta\hat{k}}(u_2)\,
 \R_{1,2}(u_1-u_2)\no\\
 &&\qquad\quad= \sum_{i,j}R(u_1-u_2;m)^{kl}_{ij}\,
 \tilde{\phi}^1_{m+\eta(\hat{\imath}+\hat{\jmath}),m+\eta\hat{\jmath}}(u_1)
 \tilde{\phi}^2_{m+\eta\hat{\jmath},m}(u_2),\label{Face-vertex2}\\
 &&\bar{\phi}^2_{m,m-\eta\hat{l}}(u_2)\,\R_{1,2}(u_1-u_2)
   \phi^1_{m,m-\eta\hat{\imath}}(u_1)\no\\
 &&\qquad\quad= \sum_{k,j}R(u_1-u_2;m)^{kl}_{ij}\,
 \phi^1_{m-\eta\hat{l},m-\eta(\hat{k}+\hat{l})}(u_1)
 \bar{\phi}^2_{m-\eta\hat{\imath},m-\eta(\hat{\imath}+\hat{\jmath})}(u_2),\label{Face-vertex3}\\
 &&\bar{\phi}^1_{m-\eta\hat{l},m-\eta(\hat{k}+\hat{l})}(u_1)
 \bar{\phi}^2_{m,m-\eta\hat{l}}(u_2)\,\R_{12}(u_1-u_2)\no\\
 &&\qquad\quad= \sum_{i,j}R(u_1-u_2;m)^{kl}_{ij}\,
 \bar{\phi}^1_{m,m-\eta\hat{\imath}}(u_1)
 \bar{\phi}^2_{m-\eta\hat{\imath},m-\eta(\hat{\imath}
 +\hat{\jmath})}(u_2).\label{Face-vertex4}
\end{eqnarray}

In addition to the Riemann identity (\ref{identity}), the $\s$-function enjoys the
following properties:
\bea
 &&\s(2u)=\frac{2\s(u)\,\s_{(0,1)}(u)\,\s_{(1,0)}(u)\,\s_{(1,1)}(u)}
     {\s_{(0,1)}(0)\,\s_{(1,0)}(0)\,\s_{(1,1)}(0)},\\
 &&\s(u+1)=-\s(u),\quad\quad \s(u+\tau)=e^{-2i\pi(u+\frac{1}{2}+\frac{\tau}{2})}\s(u),
\eea where the functions $\s_{a}(u)$ are given by (\ref{sigma-function}).
Using the above identities and  the method in \cite{Fan98}, after tedious calculations,
we can show that the K-matrices $K^{\pm}(u)$ given by
(\ref{K-matrix}) and (\ref{DK-matrix}) can be expressed in terms
of the intertwiners and {\it diagonal\/} matrices $\K(\l|u)$ and
$\tilde{\K}(\l|u)$ as follows \bea &&K^-(u)^s_t=
\sum_{i,j}\phi^{(s)}_{\l-\eta(\hat{\imath}-\hat{\jmath}),
~\l-\eta\hat{\imath}}(u)
\K(\l|u)^j_i\bar{\phi}^{(t)}_{\l,~\l-\eta\hat{\imath}}(-u),\label{K-F-1}\\
&&K^+(u)^s_t= \sum_{i,j}
\phi^{(s)}_{\l,~\l-\eta\hat{\jmath}}(-u)\tilde{\K}(\l|u)^j_i
\tilde{\phi}^{(t)}_{\l-\eta(\hat{\jmath}-\hat{\imath}),
~\l-\eta\hat{\jmath}}(u).\label{K-F-2}\eea Here the two {\it
diagonal\/} matrices $\K(\l|u)$ and $\tilde{\K}(\l|u)$ are given
by \bea
&&\K(\l|u)\equiv{\rm Diag}(k(\l|u)_1,\,k(\l|u)_2)={\rm
Diag}(\frac{\s(\l_1+\xi-u)}{\s(\l_1+\xi+u)},\,
\frac{\s(\l_2+\xi-u)}{\s(\l_2+\xi+u)}),\label{K-F-3}\\
&&\tilde{\K}(\l|u)\equiv{\rm
Diag}(\tilde{k}(\l|u)_1,\,\tilde{k}(\l|u)_2)\no\\
&&~~~~~~~~~={\rm
Diag}(\frac{\s(\l_{12}\hspace{-0.1cm}-\hspace{-0.1cm}
\eta)\s(\l_1\hspace{-0.1cm}+\hspace{-0.1cm}\bar{\xi}+\hspace{-0.1cm}u
\hspace{-0.1cm}+\hspace{-0.1cm}\eta)}
{\s(\l_{12})\s(\l_1+\bar{\xi}-u-\eta)},\,
\frac{\s(\l_{12}\hspace{-0.1cm}+\hspace{-0.1cm}
\eta)\s(\l_2\hspace{-0.1cm}+\hspace{-0.1cm}\bar{\xi}\hspace{-0.1cm}
+\hspace{-0.1cm}u\hspace{-0.1cm}+\hspace{-0.1cm}\eta)}
{\s(\l_{12})\s(\l_2+\bar{\xi}-u-\eta)}).\label{K-F-4} \eea
Although the vertex type K-matrices $K^{\pm}(u)$ given by
(\ref{K-matrix}) and (\ref{DK-matrix}) are generally non-diagonal,
after the face-vertex transformations (\ref{K-F-1}) and
(\ref{K-F-2}), the face type counterparts $\K(\l|u)$ and
$\tilde{\K}(\l|u)$  become {\it simultaneously\/} diagonal. This
fact enabled the authors in \cite{Yan04-1,Yan07} to diagonalize the transfer matrices $\t(u)$ (\ref{trans})
by applying the generalized algebraic Bethe
ansatz method developed in \cite{Yan04}.
.

\subsection{Two sets of eigenstates}

In order to construct the Bethe states of the open XYZ model
with non-diagonal boundary terms specified by the K-matrices
(\ref{K-matrix-2-1}) and (\ref{K-matrix-6}), we need to introduce
the new double-row monodromy matrices $\T^{\pm}(m|u)$ \cite{Yan04,Yan10,Fen10}:
\bea
 \T^-(m|u)^{\nu}_{\mu}
     &=&\tilde{\phi}^{0}_{m-\eta(\hat{\mu}-\hat{\nu}),
        m-\eta\hat{\mu}}(u)~\mathbb{T}_0(u)\phi^{0}_{m,
        m-\eta\hat{\mu}}(-u),\label{Mon-F}\\
\T^+(m|u)^j_i
     &=&\prod_{k\neq j}\frac{\s(m_{jk})}{\s(m_{jk}-\eta)}
        \,\phi^{t_0}_{m-\eta(\hat{\jmath}-\hat{\imath}),m-\eta\hat{\jmath}}(u)
        \lt(\mathbb{T}^+(u)\rt)^{t_0}\bar{\phi}^{t_0}_{m,m-\eta\hat{\jmath}}(-u),
        \label{Mon-F-1}
\eea where $t_0$ denotes transposition in the $0$-th space (i.e.
auxiliary space) and $\mathbb{T}^+(u)$ is given by
\bea
  \lt(\mathbb{T}^+(u)\rt)^{t_0}&=&T^{t_0}(u)\lt(K^+(u)\rt)^{t_0}\hat{T}^{t_0}(u).
      \label{Mon-V-1}
\eea
These double-row monodromy matrices, in the face picture, can be
expressed in terms of the face type R-matrix $R(u;m)$
(\ref{R-matrix}) and  K-matrices  $\K(\l|u)$ (\ref{K-F-3}) and $\tilde{\K}(\l|u)$
(\ref{K-F-4}) (for the details see section 3).

So far  only two sets of Bethe states ( i.e. eigenstates) of the transfer matrix
for the models with non-diagonal boundary terms  have been found \cite{Fan96,Yan07,Fen10}.
These two sets of states are
\bea
&&|\{v^{(1)}_i\}\rangle^{(I)}=
     \T^+(\l+2\eta\hat{1}|v^{(1)}_1)^1_2\cdots
   \T^+(\l+2M\eta\hat{1}|v^{(1)}_M)^1_2|\O^{(I)}(\l)\rangle,
   \label{Bethe-state-1}\\
&&|\{v^{(2)}_i\}\rangle^{(II)} =
   \T^-(\l-2\eta\hat{2}|v^{(2)}_1)^2_1
   \cdots
   \T^-(\l\hspace{-0.04truecm}-\hspace{-0.04truecm}2M\eta\hat{2}|v^{(2)}_M)^2_1|\O^{(II)}(\l)\rangle,
   \label{Bethe-state-2}
\eea where the vector $\l$ is related to the boundary parameters
(\ref{boundary-vector}). The associated reference states
$|\O^{(I)}(\l)\rangle$ and $|\O^{(II)}(\l)\rangle$ are \bea
\hspace{-1.2truecm}|\O^{(I)}(\l)\rangle
   &=&\phi^1_{\l+N\eta\hat{1},\l+(N-1)\eta\hat{1}}(z_1)
      \phi^2_{\l+(N-1)\eta\hat{1},\l+(N-2)\eta\hat{1}}(z_{2})\cdots
      \phi^N_{\l+\eta\hat{1},\l}(z_N),\label{Vac-1}\\
\hspace{-1.2truecm} |\O^{(II)}(\l)\rangle&=&
\phi^1_{\l,\l-\eta\hat{2}}(z_1)
\phi^{2}_{\l-\eta\hat{2},\l-2\eta\hat{2}}(z_{2})\cdots
\phi^N_{\l-(N-1)\eta\hat{2},\l-N\eta\hat{2}}(z_N).\label{Vac-2}
\eea It is remarked that   $\phi^k={\rm id}\otimes {\rm
id}\cdots\otimes \stackrel{k-th}{\phi}\otimes {\rm id}\cdots$.

If the parameters $\{v^{(1)}_k\}$ satisfy the first set of  Bethe
ansatz equations given by
\bea &&\hspace{-0.1cm}\frac
{\s(\l_2+\xi+v^{(1)}_{\a})\s(\l_2+\bar\xi-v^{(1)}_{\a})
\s(\l_1+\bar\xi+v^{(1)}_{\a})\s(\l_1+\xi-v^{(1)}_{\a})}
{\s(\l_2\hspace{-0.1cm}+\hspace{-0.1cm}
\bar\xi\hspace{-0.1cm}+\hspace{-0.1cm}v^{(1)}_{\a}
\hspace{-0.1cm}+\hspace{-0.1cm}\eta)
\s(\l_2\hspace{-0.1cm}+\hspace{-0.1cm}\xi\hspace{-0.1cm}-\hspace{-0.1cm}v^{(1)}_{\a}
\hspace{-0.1cm}-\hspace{-0.1cm}\eta)
\s(\l_1\hspace{-0.1cm}+\hspace{-0.1cm}\xi\hspace{-0.1cm}+\hspace{-0.1cm}
v^{(1)}_{\a}\hspace{-0.1cm}+\hspace{-0.1cm}\eta)
\s(\l_1\hspace{-0.1cm}+\hspace{-0.1cm}\bar\xi\hspace{-0.1cm}-\hspace{-0.1cm}v^{(1)}_{\a}
\hspace{-0.1cm}-\hspace{-0.1cm}\eta)}\no\\
&&~~~~~~=\prod_{k\neq
\a}^M\frac{\s(v^{(1)}_{\a}+v^{(1)}_k+2\eta)\s(v^{(1)}_{\a}-v^{(1)}_k+\eta)}
{\s(v^{(1)}_{\a}+v^{(1)}_k)\s(v^{(1)}_{\a}-v^{(1)}_k-\eta)}\no\\
&&~~~~~~~~~~\times\prod_{k=1}^{2M}\frac{\s(v^{(1)}_{\a}+z_k)\s(v^{(1)}_{\a}-z_k)}
{\s(v^{(1)}_{\a}+z_k+\eta)\s(v^{(1)}_{\a}-z_k+\eta)},~~\a=1,\cdots,M,
\label{BA-D-1}\eea the Bethe state
$|v^{(1)}_1,\cdots,v^{(1)}_M\rangle^{(1)}$ becomes the eigenstate
of the transfer matrix with eigenvalue $\L^{(1)}(u)$  given by
\cite{Fen10}
\bea
&&\L^{(1)}(u)=\frac{\s(\l_2+\bar\xi-u)\s(\l_1+\bar\xi+u)\s(\l_1+\xi-u)\s(2u+2\eta)}
{\s(\l_2+\bar\xi-u-\eta)\s(\l_1+\bar\xi-u-\eta)\s(\l_1+\xi+u)\s(2u+\eta)}\no\\
&&~~~~~~~~~~~~~~~~~~\times\prod_{k=1}^M\frac{\s(u+v^{(1)}_k)\s(u-v^{(1)}_k-\eta)}
{\s(u+v^{(1)}_k+\eta)\s(u-v^{(1)}_k)}\no\\
&&~~~~~~+\frac{\s(\l_2+\bar\xi+u+\eta)\s(\l_1+\xi+u+\eta)\s(\l_2+\xi-u-\eta)\s(2u)}
{\s(\l_2+\bar\xi-u-\eta)\s(\l_1+\xi+u)\s(\l_2+\xi+u)\s(2u+\eta)}\no\\
&&~~~~~~~~~~~~~~~~~~\times\prod_{k=1}^M\frac{\s(u+v^{(1)}_k+2\eta)\s(u-v^{(1)}_k+\eta)}
{\s(u+v^{(1)}_k+\eta)\s(u-v^{(1)}_k)}\no\\
&&~~~~~~~~~~~~~~~~~~\times\prod_{k=1}^{2M}\frac{\s(u+z_k)\s(u-z_k)}
{\s(u+z_k+\eta)\s(u-z_k+\eta)}.\label{Eigenfuction-D-1}
 \eea

\noindent If the parameters $\{v^{(2)}_k\}$ satisfy the second
Bethe Ansatz equations
\bea
&&\hspace{-0.1cm}\frac
  {\s(\l_1+\xi+v^{(2)}_{\a})\s(\l_1+\bar\xi-v^{(2)}_{\a})
  \s(\l_2+\bar\xi+v^{(2)}_{\a})\s(\l_2+\xi-v^{(2)}_{\a})}
  {\s(\l_1\hspace{-0.1cm}+\hspace{-0.1cm}
  \bar\xi\hspace{-0.1cm}+\hspace{-0.1cm}v^{(2)}_{\a}
  \hspace{-0.1cm}+\hspace{-0.1cm}\eta)
  \s(\l_1\hspace{-0.1cm}+\hspace{-0.1cm}\xi\hspace{-0.1cm}-\hspace{-0.1cm}v^{(2)}_{\a}
  \hspace{-0.1cm}-\hspace{-0.1cm}\eta)
  \s(\l_2\hspace{-0.1cm}+\hspace{-0.1cm}\xi\hspace{-0.1cm}+\hspace{-0.1cm}
  v^{(2)}_{\a}\hspace{-0.1cm}+\hspace{-0.1cm}\eta)
  \s(\l_2\hspace{-0.1cm}+\hspace{-0.1cm}\bar\xi\hspace{-0.1cm}-\hspace{-0.1cm}v^{(2)}_{\a}
  \hspace{-0.1cm}-\hspace{-0.1cm}\eta)}\no\\
&&~~~~~~=\prod_{k\neq
  \a}^M\frac{\s(v^{(2)}_{\a}+v^{(2)}_k+2\eta)\s(v^{(2)}_{\a}-v^{(2)}_k+\eta)}
  {\s(v^{(2)}_{\a}+v^{(2)}_k)\s(v^{(2)}_{\a}-v^{(2)}_k-\eta)}\no\\
&&~~~~~~~~~~\times\prod_{k=1}^{2M}\frac{\s(v^{(2)}_{\a}+z_k)\s(v^{(2)}_{\a}-z_k)}
  {\s(v^{(2)}_{\a}+z_k+\eta)\s(v^{(2)}_{\a}-z_k+\eta)},~~\a=1,\cdots,M,
  \label{BA-D-2}
\eea the Bethe states $|v^{(2)}_1,\cdots,v^{(2)}_M\rangle^{(II)}$
yield the second set of the eigenstates of the transfer matrix
with the eigenvalues \cite{Fan96,Yan07},
\bea
&&\L^{(2)}(u)=\frac{\s(2u+2\eta)\s(\l_1+\bar\xi-u)\s(\l_2+\bar\xi+u)\s(\l_2+\xi-u)}
{\s(2u+\eta)\s(\l_1+\bar\xi-u-\eta)\s(\l_2+\bar\xi-u-\eta)\s(\l_2+\xi+u)}\no\\
&&~~~~~~~~~~~~~~~~~~\times\prod_{k=1}^M\frac{\s(u+v^{(2)}_k)\s(u-v^{(2)}_k-\eta)}
{\s(u+v^{(2)}_k+\eta)\s(u-v^{(2)}_k)}\no\\
&&~~~~~~+\frac{\s(2u)\s(\l_1+\bar\xi+u+\eta)
\s(\l_2+\xi+u+\eta)\s(\l_1+\xi-u-\eta)}
{\s(2u+\eta)\s(\l_1+\bar\xi-u-\eta)\s(\l_2+\xi+u)\s(\l_1+\xi+u)}\no\\
&&~~~~~~~~~~~~~~~~~~\times\prod_{k=1}^M\frac{\s(u+v^{(2)}_k+2\eta)\s(u-v^{(2)}_k+\eta)}
{\s(u+v^{(2)}_k+\eta)\s(u-v^{(2)}_k)}\no\\
&&~~~~~~~~~~~~~~~~~~\times\prod_{k=1}^{2M}\frac{\s(u+z_k)\s(u-z_k)}
{\s(u+z_k+\eta)\s(u-z_k+\eta)}.\label{Eigenfuction-D-2}
 \eea


\section{ Scalar products}
\label{T} \setcounter{equation}{0}

It was shown that in order to compute correlation functions of
the closed chain \cite{Kor93} and the open chain with
diagonal boundary terms \cite{Wan02,Kit07},
one suffices to calculate the scalar products
of an  on-shell Bethe state and a general state  (an off-shell Bethe state).
The aim of this paper is to give the explicit expressions of the following
scalar products  of the open XYZ chain with non-diagonal
boundary terms:
\bea
&&\hspace{-1.2truecm}S^{I,II}(\{u_{\a}\};\{v^{(2)}_i\})=
   {}^{(I)}\langle\{u_{\a}\}|\{v^{(2)}_i\}\rangle^{(II)},\quad
   S^{II,I}(\{u_{\a}\};\{v^{(1)}_i\})= {}^{(II)}\langle\{u_{\a}\}
   |\{v^{(1)}_i\}\rangle^{(I)},\label{Scalar-1}\\
&&\hspace{-1.2truecm}S^{I,I}(\{u_{\a}\};\{v^{(1)}_i\})=
   {}^{(I)}\langle\{u_{\a}\}|\{v^{(1)}_i\}\rangle^{(I)},\quad
   S^{II,II}(\{u_{\a}\};\{v^{(2)}_i\})=
   {}^{(II)}\langle\{u_{\a}\}|\{v^{(2)}_i\}\rangle^{(II)},\label{Scalar-2}
\eea where the dual states ${}^{(I)}\langle\{u_{\a}\}|$ and
${}^{(II)}\langle\{u_{\a}\}|$ are given by
\bea
&&{}^{(I)}\langle\{u_{\a}\}|=\langle\Omega^{(I)}(\l)|
   \T^-(\l-2(M-1)\eta\hat{1}|u_M)^2_1\ldots\T^-(\l|u_1)^2_1,\label{Dual-1}\\
&&{}^{(II)}\langle\{u_{\a}\}|=\langle\Omega^{(II)}(\l)|
   \T^+(\l+2(M-1)\eta\hat{2}|u_M)^1_2\ldots\T^+(\l|u_1)^1_2,\label{Dual-2}
\eea  and $\langle\Omega^{(I)}(\l)|$, $\langle\Omega^{(II)}(\l)|$ are
\bea
 && \langle\Omega^{(I)}(\l)|=\tilde{\phi}^1_{\l,\l-\eta\hat{1}}(z_1)\ldots
    \tilde{\phi}^N_{\l-(2M-1)\eta\hat{1},\l-2M\eta\hat{1}}(z_{N}),\\
 && \langle\Omega^{(II)}(\l)|=\tilde{\phi}^1_{\l+2M\eta\hat{2},\l+(2M-1)\eta\hat{2}}(z_1)\ldots
    \tilde{\phi}^N_{\l+\eta\hat{1},\l}(z_{N}).
\eea  Some remarks are in order. The parameters $\{u_{\a}\}$ in (\ref{Dual-1})-(\ref{Dual-2})
are free parameters, namely, they do not
need to satisfy the Bethe ansatz equations. Moreover, even if these parameters are required satisfy the associated
Bethe ansatz equations (\ref{BA-D-1}) and (\ref{BA-D-2})
respectively, the corresponding  dual states (\ref{Dual-1}) and (\ref{Dual-2}) are not the eigenstates of
the model (in contrast with those of the open XXZ chain with diagonal boundary terms \cite{Kit07}). Such a phenomena
was already found for the open XXZ chain with non-diagonal boundary terms \cite{Yan11-1}.

The K-matrices $K^{\pm}(u)$ given by (\ref{K-matrix}) and
(\ref{DK-matrix}) are generally non-diagonal (in the vertex
picture), after the face-vertex transformations (\ref{K-F-1}) and
(\ref{K-F-2}), the face type counterparts $\K(\l|u)$ and
$\tilde{\K}(\l|u)$ given by (\ref{K-F-3}) and (\ref{K-F-4}) {\it
simultaneously\/} become diagonal. This fact suggests that it
would be much simpler if one performs all calculations in the face
picture.
\subsection{Face picture}

Let us introduce the face type one-row monodromy matrix (c.f
(\ref{Mon-V})) \bea
 T_{F}(l|u)&\equiv &T^{F}_{0,1\ldots N}(l|u)\no\\
 &=&R_{0,N}(u-z_N;l-\eta\sum_{i=1}^{N-1}h^{(i)})\ldots
    R_{0,2}(u-z_2;l-\eta h^{(1)})R_{0,1}(u-z_1;l),\no\\
 &=&\lt(\begin{array}{ll}T_F(l|u)^1_1&T_F(l|u)^1_2\\T_F(l|u)^2_1&
   T_F(l|u)^2_2\end{array}\rt)
    \label{Monodromy-face-1}
\eea where $l$ is a generic vector in $V$. The monodromy matrix
satisfies the face type quadratic exchange relation
\cite{Fel96,Hou03}. Applying $T_F(l|u)^i_j$ to an arbitrary vector
$|i_1,\ldots,i_N\rangle$ in the N-tensor product space $V^{\otimes
N}$ given by \bea
   |i_1,\ldots,i_N\rangle=\e^1_{i_1}\ldots
   \e^N_{i_N},\label{Vector-V}
\eea we have \bea
 T_F(l|u)^i_j|i_1,\ldots,i_N\rangle&\equiv&
    T_F(m;l|u)^i_j|i_1,\ldots,i_N\rangle\no\\
 &=&\sum_{\a_{N-1}\ldots\a_1}\sum_{i'_N\ldots i'_1}
 R(u-z_N;l-\eta\sum_{k=1}^{N-1}\hat{\imath}'_k)
   ^{i\,\,\,\,\,\,\,\,\,\,\,\,\,\,i'_N}_{\a_{N-1}\,i_N}\ldots\no\\
 &&\quad\quad\times R(u-z_2;l-\eta\hat{\imath}'_1)^{\a_2\,i'_2}_{\a_1\,\,i_2}
 R(u-z_1;l)^{\a_1\,i'_1}_{j\,\,\,\,i_1}
   \,\,|i'_1,\ldots,i'_N\rangle,\label{Monodromy-face-2}
\eea where $m=l-\eta\sum_{k=1}^N\hat{\imath}_k$. With the help of the crossing
relation (\ref{Crossing}), the face-vertex
correspondence relation (\ref{Face-vertex}) and the relations (\ref{Int2}),
following the
method developed in \cite{Yan04,Yan10,Yan11-2}, we find that the scalar products (\ref{Scalar-1})-(\ref{Scalar-2})
can be expressed in terms of the face-type double-row monodromy operators as follows:
\bea
 &&S^{I,II}(\{u_{\a}\};\{v^{(2)}_i\})=\langle 1,\ldots,1|\T^-_F(\l-2(M-1)\eta\hat{1},\l|u_M)^2_1
     \ldots\T^-_F(\l,\l|u_1)^2_1 \no\\
     &&\qquad\qquad\times \T^-_F(\l+2\eta\hat{1},\l|v^{(2)}_1)^2_1 \ldots
     \T^-_F(\l+2M\eta\hat{1},\l|v^{(2)}_M)^2_1 |2,\ldots,2\rangle ,\label{Scalar-3}\\
 &&S^{II,I}(\{u_{\a}\};\{v^{(1)}_i\})=\langle 2,\ldots,2|\T^+_F(\l,\l+2(M-1)\eta\hat{2}|u_M)^1_2
     \ldots\T^+_F(\l,\l|u_1)^1_2 \no\\
     &&\qquad\qquad\times \T^+_F(\l,\l-2\eta\hat{2}|v^{(1)}_1)^1_2 \ldots
     \T^+_F(\l,\l-2M\eta\hat{2}|v^{(1)}_M)^1_2|1,\ldots,1\rangle,\\
 &&S^{I,I}(\{u_{\a}\};\{v^{(1)}_i\})=\langle 1,\ldots,1|\T^-_F(\l-2(M-1)\eta\hat{1},\l|u_M)^2_1
     \ldots\T^-_F(\l,\l|u_1)^2_1 \no\\
     &&\qquad\qquad\times \T^+_F(\l,\l+2\eta\hat{1}|v^{(1)}_1)^1_2 \ldots
     \T^+_F(\l,\l+2M\eta\hat{1}|v^{(1)}_M)^1_2|1,\ldots,1\rangle,\\
 &&S^{II,II}(\{u_{\a}\};\{v^{(2)}_i\})=\langle 2,\ldots,2|\T^+_F(\l,\l+2(M-1)\eta\hat{2}|u_M)^1_2
     \ldots\T^+_F(\l,\l|u_1)^1_2 \no\\
     &&\qquad\qquad\times \T^-_F(\l-2\eta\hat{2},\l|v^{(2)}_1)^2_1 \ldots
     \T^-_F(\l-2M\eta\hat{2},\l|v^{(2)}_M)^2_1 |2,\ldots,2\rangle.\label{Scalar-4}
\eea  The above double-row monodromy  matrix operators $\T^-_F(m,\l|u)^2_1$ and  $\T^+_F(\l,m|u)^1_2$ are given
in terms of the one-row monodromy matrix operator $T_F(m;l|u)^i_j$
\cite{Yan11-2}
\bea
 &&\T^-_F(m,\l|u)^2_1=
 \frac{\s(m_{21})}{\s(\l_{21})}\prod_{k=1}^N
      \frac{\s(u+z_k)}{\s(u+z_k+\eta)}\no\\
 &&\,\quad\times\lt\{
      \frac{\s(\l_1+\xi-u)}{\s(\l_1+\xi+u)}
      T_F(m,\l|u)^2_1
      T_F(m+\eta\hat{2},\l+\eta\hat{2}|-u-\eta)^2_2\rt.\no\\
 &&\,\qquad-\lt.
      \frac{\s(\l_2+\xi-u)}{\s(\l_2+\xi+u)}
      T_F(m+2\eta\hat{2},\l|u)^2_2
      T_F(m+\eta\hat{1},\l+\eta\hat{1}|-u-\eta)^2_1\rt\},\label{Expression-3}\\
 &&\T^+_F(\l,m|u)^1_2=\prod_{k=1}^N
      \frac{\s(u+z_k)}{\s(u+z_k+\eta)}\no\\
 &&\,\quad\times\lt\{
      \frac{\s(\l_{12}\hspace{-0.08truecm}-\hspace{-0.08truecm}\eta)
      \s(\l_1\hspace{-0.08truecm}+\hspace{-0.08truecm}\bar{\xi}
      \hspace{-0.08truecm}+\hspace{-0.08truecm}u\hspace{-0.08truecm}+\hspace{-0.08truecm}\eta)}
      {\s(m_{12}\hspace{-0.08truecm}-\hspace{-0.08truecm}\eta)
      \s(\l_1\hspace{-0.08truecm}+\hspace{-0.08truecm}\bar{\xi}\hspace{-0.08truecm}-
      \hspace{-0.08truecm}u\hspace{-0.08truecm}-\hspace{-0.08truecm}\eta)}
      T_F(\l\hspace{-0.08truecm}+\hspace{-0.08truecm}2\eta\hat{2},m
      \hspace{-0.08truecm}+\hspace{-0.08truecm}2\eta\hat{2}|u)^1_2
      T_F(\l\hspace{-0.08truecm}+\hspace{-0.08truecm}\eta\hat{2},m
      \hspace{-0.08truecm}+\hspace{-0.08truecm}\eta\hat{2}|
      \hspace{-0.08truecm}-\hspace{-0.08truecm}u
      \hspace{-0.08truecm}-\hspace{-0.08truecm}\eta)^2_2\rt.\no\\
 &&\,\qquad-\lt.
      \frac{\s(\l_{21}\hspace{-0.08truecm}-\hspace{-0.08truecm}\eta)
      \s(\l_2\hspace{-0.08truecm}+\hspace{-0.08truecm}\bar{\xi}
      \hspace{-0.08truecm}+\hspace{-0.08truecm}u
      \hspace{-0.08truecm}+\hspace{-0.08truecm}\eta)}
      {\s(m_{21}\hspace{-0.08truecm}+\hspace{-0.08truecm}\eta)
      \s(\l_2\hspace{-0.08truecm}+\hspace{-0.08truecm}\bar{\xi}
      \hspace{-0.08truecm}-\hspace{-0.08truecm}u
      \hspace{-0.08truecm}-\hspace{-0.08truecm}\eta)}
      T_F(\l,m\hspace{-0.08truecm}+\hspace{-0.08truecm}2\eta\hat{2}|u)^2_2
      T_F(\l\hspace{-0.08truecm}+\hspace{-0.08truecm}\eta\hat{2},m
      \hspace{-0.08truecm}+\hspace{-0.08truecm}\eta\hat{2}|
      \hspace{-0.08truecm}-\hspace{-0.08truecm}u\hspace{-0.08truecm}-\hspace{-0.08truecm}
      \eta)^1_2\rt\}.\no\\
 &&\label{Expression-4}
\eea

In the next section we use the Drinfeld twist (or
factorizing F-matrix) of the eight-vertex SOS model proposed in \cite{Alb00} to construct
the polarization free forms of the two sets of
pseudo-particle creation/annihilation operators $\T^{\pm}_F$ given by
(\ref{Expression-3}) and (\ref{Expression-4}), which
allow us to construct the explicit expressions of the scalar products (\ref{Scalar-3})-
(\ref{Scalar-4}).


\section{ F-basis}
\label{F} \setcounter{equation}{0}

In this section, after briefly reviewing the result \cite{Alb00} about the Drinfeld twist
\cite{Dri83} (factorizing F-matrix) of the eight-vertex SOS model, we
obtain the explicit expression of the double rows
monodromy operator $\T^{\mp}_F(m,\l|u)^2_1$ given by (\ref{Expression-3}) and (\ref{Expression-4}) in
the F-basis provided by the F-matrix.

\subsection{Factorizing Drinfeld twist $F$}
Let $ \mathcal{S}_N$ be the permutation group over indices
$1,\ldots,N$ and $\{s_i|i=1,\ldots,N-1\}$ be the set of
elementary permutations in $\mathcal{S}_N$. For each elementary
permutation $s_i$, we introduce the associated operator
$R^{s_i}_{1\ldots N}$ on the quantum space
\bea
  R^{s_i}_{1\ldots N}(l)\equiv R^{s_i}(l)=R_{i,i+1}
    (z_i-z_{i+1}|l-\eta\sum_{k=1}^{i-1}h^{(k)}),\label{Fundamental-R-operator}
\eea where $l$ is a generic vector in $V$. For any $s,\,s'\in
\mathcal{S}_N$, operator $R^{ss'}_{1\ldots N}$ associated with
$ss'$ satisfies the following composition law
 \cite{Mai00,Alb00,Yan04-2,Zha06-1}:
\bea
  R_{1\ldots N}^{ss'}(l)=R^{s'}_{s(1\ldots
  N)}(l)\,R^{s}_{1\ldots N}(l).\label{Rule}
\eea Let $s$ be decomposed in a minimal way in terms of
elementary permutations,
\bea
  s=s_{\b_1}\ldots s_{\b_p}, \label{decomposition}
\eea where $\b_i=1,\ldots, N-1$ and the positive integer $p$ is
the length of $s$. The composition law (\ref{Rule}) enables one
to obtain  operator $R^{s}_{1\ldots N}$ associated with each
$s\in\mathcal{S}_N $. The dynamical quantum Yang-Baxter equation
(\ref{MYBE}), the weight conservation condition (\ref{Conservation})
and the unitary condition (\ref{Unitary}) guarantee the uniqueness of
$R^{s}_{1\ldots N}$. Moreover, one may check that
$R^{s}_{1\ldots N}$ satisfies the following exchange relation
with the face type one-row monodromy matrix
(\ref{Monodromy-face-1}) \bea
  R^{s}_{1\ldots N}(l)T^F_{0,1\ldots N}(l|u)=T^F_{0,s(1\ldots N)}(l|u)
    R^{s}_{1\ldots N}(l-\eta h^{(0)}),\quad\quad \forall s\in
    \mathcal{S}_N.\label{Exchang-Face-1}
\eea

Now, we construct the face-type Drinfeld twist $F_{1\ldots
N}(l)\equiv F_{1\ldots N}(l;z_1,\ldots,z_N)$ \footnote{In this
paper, we adopt the convention: $F_{s(1\ldots N)}(l)\equiv
F_{s(1\ldots N)}(l;z_{s(1)},\ldots,z_{s(N)})$.} on the $N$-fold
tensor product space $V^{\otimes N}$, which  satisfies the
following three properties \cite{Alb00,Yan04-2,Zha06-1}:
\bea
 &&{\rm I.\,\,\,\,lower-triangularity;}\\
 &&{\rm II.\,\,\, non-degeneracy;}\\
 &&{\rm III.\,factorizing \, property}:\,\,
 R^{s}_{1\ldots N}(l)\hspace{-0.08truecm}=\hspace{-0.08truecm}
    F^{-1}_{s(1\ldots N)}(l)F_{1\ldots N}(l), \,\,
 \forall s\in  \mathcal{S}_N.\label{Factorizing}
\eea Substituting (\ref{Factorizing}) into the exchange relation
(\ref{Exchang-Face-1}), we have
\bea
 F^{-1}_{s(1\ldots N)}(l)F_{1\ldots N}(l)T^F_{0,1\ldots N}(l|u)=
   T^F_{0,s(1\ldots N)}(l|u)F^{-1}_{s(1\ldots N)}(l-\eta h^{(0)})
   F_{1\ldots N}(l-\eta h^{(0)}).
\eea Equivalently,
\bea
 F_{1\ldots N}(l)T^F_{0,1\ldots N}(l|u)F^{-1}_{1\ldots N}(l-\eta h^{(0)})
   =F_{s(1\ldots N)}(l)T^F_{0,s(1\ldots N)}(l|u)
   F^{-1}_{s(1\ldots N)}(l-\eta h^{(0)}).\label{Invariant}
\eea Let us introduce the twisted monodromy matrix
$\tilde{T}^F_{0,1\ldots N}(l|u)$ by \bea
 \tilde{T}^F_{0,1\ldots N}(l|u)&=&
  F_{1\ldots N}(l)T^F_{0,1\ldots N}(l|u)F^{-1}_{1\ldots N}(l-\eta
  h^{(0)})\no\\
  &=&\lt(\begin{array}{ll}\tilde{T}_F(l|u)^1_1&\tilde{T}_F(l|u)^1_2
  \\\tilde{T}_F(l|u)^2_1&
   \tilde{T}_F(l|u)^2_2\end{array}\rt).\label{Twisted-Mon-F}
\eea Then (\ref{Invariant}) implies that the twisted monodromy
matrix is symmetric under $\mathcal{S}_N$, namely, \bea
 \tilde{T}^F_{0,1\ldots N}(l|u)=\tilde{T}^F_{0,s(1\ldots
 N)}(l|u), \quad \forall s\in \mathcal{S}_N.
\eea

Define the F-matrix:
\bea
  F_{1\ldots N}(l)=\sum_{s\in
     \mathcal{S}_N}\sum^2_{\{\a_j\}=1}\hspace{-0.22truecm}{}^*\,\,\,\,
     \prod_{i=1}^NP^{s(i)}_{\a_{s(i)}}
     \,R^{s}_{1\ldots N}(l),\label{F-matrix}
\eea where $P^i_{\a}$ is the embedding of the project operator
$P_{\a}$ in the $i^{{\rm th}}$ space with matric elements
$(P_{\a})_{kl}=\d_{kl}\d_{k\a}$. The sum $\sum^*$ in
(\ref{F-matrix}) is over all non-decreasing sequences of the
labels $\a_{s(i)}$:
\bea
  && \a_{s(i+1)}\geq \a_{s(i)}\quad {\rm if}\quad s(i+1)>s(i),\\
  && \a_{s(i+1)}> \a_{s(i)}\quad {\rm if}\quad
  s(i+1)<s(i).\label{Condition}
\eea From (\ref{Condition}), $F_{1\ldots N}(l)$ obviously is a
lower-triangular matrix. Moreover, the F-matrix is non-degenerate
because  all its diagonal elements are non-zero. It was shown in \cite{Alb00} that
the F-matrix also satisfies the factorizing property (\ref{Factorizing}).

\subsection{Completely symmetric  representations}
In the F-basis provided by the F-matrix (\ref{F-matrix}),  the
twisted operators $\tilde{T}_F(l|u)^j_i$ defined by
(\ref{Twisted-Mon-F}) become polarization free \cite{Alb00}. Here we present the
results relevant for our purpose
\bea
 &&\tilde{T}_F(l|u)^2_2=\frac{\s(l_{21}-\eta)}{\s\lt(l_{21}-\eta+
     \eta\langle H,\e_1\rangle\rt)}\otimes_{i}
     \lt(\begin{array}{ll}\frac{\s(u-z_i)}{\s(u-z_i+\eta)}&\\
     &1\end{array}\rt)_{(i)},\\[6pt]
 &&\tilde{T}_F(l|u)^2_1=\sum_{i=1}^N\frac{\s(\eta)
     \s(u\hspace{-0.08truecm}-\hspace{-0.08truecm}z_i
     \hspace{-0.08truecm}+\hspace{-0.08truecm}l_{12})}
     {\s(u\hspace{-0.08truecm}-\hspace{-0.08truecm}z_i
     \hspace{-0.08truecm}+\hspace{-0.08truecm}\eta)\s(l_{12})} E_{12}^i\otimes_{j\neq i}
     \lt(\begin{array}{ll}\frac{\s(u-z_j)\s(z_i-z_j+\eta)}{\s(u-z_j+\eta)\s(z_i-z_j)}&\\
     &1\end{array}\rt)_{(j)},\\[6pt]
 &&\tilde{T}_F(l|u)^1_2=\frac{\s(l_{21}\hspace{-0.08truecm}-\hspace{-0.08truecm}\eta)}
     {\s(l_{21}\hspace{-0.08truecm}+\hspace{-0.08truecm}
     \eta\langle H,\e_1\hspace{-0.08truecm}-\hspace{-0.08truecm}\e_2\rangle)}
     \hspace{-0.08truecm}
     \sum_{i=1}^N\frac{\s(\eta)\s(u\hspace{-0.08truecm}-\hspace{-0.08truecm}z_i
     \hspace{-0.08truecm}+\hspace{-0.08truecm}l_{21}\hspace{-0.08truecm}
     +\hspace{-0.08truecm}\eta\hspace{-0.08truecm}
     +\hspace{-0.08truecm}\eta\langle H,\e_1
     \hspace{-0.08truecm}-\hspace{-0.08truecm}\e_2\rangle)}
     {\s(u\hspace{-0.08truecm}-\hspace{-0.08truecm}z_i\hspace{-0.08truecm}+\hspace{-0.08truecm}\eta)
     \s(l_{21}\hspace{-0.08truecm}+\hspace{-0.08truecm}\eta
     \hspace{-0.08truecm}+\hspace{-0.08truecm}\eta\langle
     H,\e_1\hspace{-0.08truecm}-\hspace{-0.08truecm}\e_2\rangle)}\no\\
 &&\quad\quad\quad\quad\quad\quad
     \times E_{21}^i\otimes_{j\neq i} \lt( \begin{array}{ll}
     \frac{\s(u-z_j)}{\s(u-z_j+\eta)}&\\
     &\frac{\s(z_j-z_i+\eta)}{\s(z_j-z_i)}\end{array}
     \rt)_{(j)},
\eea where $H=\sum_{k=1}^N h^{(k)}$. Applying  the above operators
to the arbitrary  state $|i_1,\ldots,i_N\rangle$ given by
(\ref{Vector-V}), we have
\bea
 &&\tilde{T}_F(m,l|u)^2_2=\frac{\s(l_{21}-\eta)}{\s\lt(l_{2}-m_1-\eta\rt)}
     \otimes_{i}
     \lt(\begin{array}{ll}\frac{\s(u-z_i)}{\s(u-z_i+\eta)}&\\
     &1\end{array}\rt)_{(i)},\\[6pt]
 &&\tilde{T}_F(m,l|u)^2_1=\sum_{i=1}^N
     \frac{\s(\eta)
     \s(u-z_i+l_{12})}{\s(u-z_i+\eta)\s(l_{12})}\no\\
 &&\quad\quad\quad\quad\quad\quad
     \times    E_{12}^i \otimes_{j\neq i}
     \lt(\begin{array}{ll}\frac{\s(u-z_j)\s(z_i-z_j+\eta)}{\s(u-z_j+\eta)\s(z_i-z_j)}&\\
     &1\end{array}\rt)_{(j)},\\[6pt]
 &&\tilde{T}_F(m,l|u)^1_2=\frac{\s(l_{21}-\eta)}
     {\s(m_{21}-2\eta)}
     \sum_{i=1}^N\frac{\s(\eta)\s(u-z_i+m_{21}-\eta)}
     {\s(u-z_i+\eta)\s(m_{21}-\eta)}\no\\
 &&\quad\quad\quad\quad\quad\quad
     \times E_{21}^i\otimes_{j\neq i} \lt( \begin{array}{ll}
     \frac{\s(u-z_j)}{\s(u-z_j+\eta)}&\\
     &\frac{\s(z_j-z_i+\eta)}{\s(z_j-z_i)}\end{array}
     \rt)_{(j)}.
\eea With the help of the Riemann identity (\ref{identity}), we find that the two pseudo-particle creation
operators (\ref{Expression-3}) and (\ref{Expression-4}) in the
F-basis simultaneously have the following completely symmetric
polarization free forms:
\bea
 &&\tilde{\T}^-_F(m,\l|u)^2_1=\frac{\s(m_{12})}{\s(m_1-\l_2)}
   \prod_{k=1}^N\frac{\s(u+z_k)}{\s(u+z_k+\eta)}\no\\
  &&\quad\quad\times \sum_{i=1}^N\frac{\s(\l_1+\xi-z_i)\s(\l_2+\xi+z_i)\s(2u) \s(\eta)}
   {\s(\l_1+\xi+u)\s(\l_2+\xi+u)\s(u-z_i+\eta)\s(u+z_i)}\no\\[6pt]
  &&\quad\quad\quad\quad\quad\quad \times
   E_{12}^i\otimes_{j\neq i}\lt(\begin{array}{ll}
   \frac{\s(u-z_j)\s(u+z_j+\eta)\s(z_i-z_j+\eta)}
   {\s(u-z_j+\eta)\s(u+z_j)\s(z_i-z_j)}&\\
   &1\end{array}\rt)_{(j)},\label{Creation-operator-1}\\[6pt]
 &&\tilde{\T}^+_F(\l,m|u)^1_2=\frac{\s(m_{21}+\eta)}{\s(m_2-\l_1)}
   \prod_{k=1}^N\frac{\s(u+z_k)}{\s(u+z_k+\eta)}\no\\
  &&\quad\quad\times \sum_{i=1}^N
   \hspace{-0.08truecm}
   \frac{\s(\l_2\hspace{-0.08truecm}+\hspace{-0.08truecm}\bar{\xi}
   \hspace{-0.08truecm}-\hspace{-0.08truecm}z_i)
   \s(\l_1\hspace{-0.08truecm}+\hspace{-0.08truecm}\bar{\xi}
   \hspace{-0.08truecm}+\hspace{-0.08truecm}z_i)
   \s(2u\hspace{-0.08truecm}+\hspace{-0.08truecm}2\eta) \s(\eta)}
   {\s(\l_1\hspace{-0.08truecm}+\hspace{-0.08truecm}\bar{\xi}
   \hspace{-0.08truecm}-\hspace{-0.08truecm}u
   \hspace{-0.08truecm}-\hspace{-0.08truecm}\eta)
   \s(\l_2\hspace{-0.08truecm}+\hspace{-0.08truecm}\bar{\xi}
   \hspace{-0.08truecm}-\hspace{-0.08truecm}u
   \hspace{-0.08truecm}-\hspace{-0.08truecm}\eta)
   \s(u\hspace{-0.08truecm}+\hspace{-0.08truecm}z_i)
   \s(u\hspace{-0.08truecm}-\hspace{-0.08truecm}z_i
   \hspace{-0.08truecm}+\hspace{-0.08truecm}\eta)}\no\\[6pt]
  &&\quad\quad\quad\quad\quad\quad \times
   E_{21}^i\otimes_{j\neq i}\lt(\begin{array}{ll}
   \frac{\s(u-z_j)\s(u+z_j+\eta)}
   {\s(u-z_j+\eta)\s(u+z_j)}&\\
   &\frac{\s(z_j-z_i+\eta)}{\s(z_j-z_i)}\end{array}\rt)_{(j)}.\label{Creation-operator-2}
\eea The very polarization free form (\ref{Creation-operator-1}) of
$\tilde{\T}^-_F(m,\l|u)^2_1$ enabled the authors in \cite{Yan11} to succeed in
obtaining  a single determinant representation of the domain wall partition function
of the eight-vertex model with a non-diagonal reflecting end.


\section{Determinant representations of the scalar products}
\label{DR} \setcounter{equation}{0}

Due to the fact that the states $|1,\ldots,1\rangle$, $|2,\ldots,2\rangle$ and their dual states
$\langle 1,\ldots,1|$, $\langle 2,\ldots,2|$ are
invariant under the action of the F-matrix $F_{1\ldots N}(l)$
(\ref{F-matrix}), the calculation of the scalar products (\ref{Scalar-3})-(\ref{Scalar-4})
can be performed in the F-basis. Namely,
\bea
 &&S^{I,II}(\{u_{\a}\};\{v^{(2)}_i\})=\langle 1,\ldots,1|\tilde{\T}^-_F(\l-2(M-1)\eta\hat{1},\l|u_M)^2_1
     \ldots\tilde{\T}^-_F(\l,\l|u_1)^2_1 \no\\
     &&\qquad\qquad\times \tilde{\T}^-_F(\l+2\eta\hat{1},\l|v^{(2)}_1)^2_1 \ldots
     \tilde{\T}^-_F(\l+2M\eta\hat{1},\l|v^{(2)}_M)^2_1 |2,\ldots,2\rangle ,\label{Scalar-5}\\
 &&S^{II,I}(\{u_{\a}\};\{v^{(1)}_i\})=\langle 2,\ldots,2|\tilde{\T}^+_F(\l,\l+2(M-1)\eta\hat{2}|u_M)^1_2
     \ldots\tilde{\T}^+_F(\l,\l|u_1)^1_2 \no\\
     &&\qquad\qquad\times \tilde{\T}^+_F(\l,\l-2\eta\hat{2}|v^{(1)}_1)^1_2 \ldots
     \tilde{\T}^+_F(\l,\l-2M\eta\hat{2}|v^{(1)}_M)^1_2|1,\ldots,1\rangle,\\
 &&S^{I,I}(\{u_{\a}\};\{v^{(1)}_i\})=\langle 1,\ldots,1|\tilde{\T}^-_F(\l-2(M-1)\eta\hat{1},\l|u_M)^2_1
     \ldots\tilde{\T}^-_F(\l,\l|u_1)^2_1 \no\\
     &&\qquad\qquad\times \tilde{\T}^+_F(\l,\l+2\eta\hat{1}|v^{(1)}_1)^1_2 \ldots
     \tilde{\T}^+_F(\l,\l+2M\eta\hat{1}|v^{(1)}_M)^1_2|1,\ldots,1\rangle,\label{Scalar-6}\\
 &&S^{II,II}(\{u_{\a}\};\{v^{(2)}_i\})=\langle 2,\ldots,2|\tilde{\T}^+_F(\l,\l+2(M-1)\eta\hat{2}|u_M)^1_2
     \ldots\tilde{\T}^+_F(\l,\l|u_1)^1_2 \no\\
     &&\qquad\qquad\times \tilde{\T}^-_F(\l-2\eta\hat{2},\l|v^{(2)}_1)^2_1 \ldots
     \tilde{\T}^-_F(\l-2M\eta\hat{2},\l|v^{(2)}_M)^2_1 |2,\ldots,2\rangle.\label{Scalar-7}
\eea In the above equations, we have used the identity: $\hat{1}=-\hat{2}$.
Thanks to the polarization free representations
(\ref{Creation-operator-1}) and (\ref{Creation-operator-2}) of the
pseudo-particle creation/annihilation  operators,  we  can obtain the determinant representations
of the scalar products.

\subsection{The scalar products $S^{I,II}$ and $S^{II,I}$}
It was shown \cite{Yan11} that the scalar product $S^{I,II}(\{u_{\a}\};\{v^{(2)}_i\})$ (resp.
$S^{II,I}(\{u_{\a}\};\{v^{(1)}_i\})$) can be expressed in terms of some determinant no matter the
parameters $\{v^{(2)}_i\}$ (resp.$\{v^{(1)}_i\}$ ) satisfy the associated Bethe ansatz equations or not. In this subsection
we do not require these parameters being the roots of the Bethe ansatz equations. Let us introduce two functions
\bea
 {\cal Z}^{(I)}_N(\{\bar{u}_{J}\})\hspace{-0.36truecm}&\equiv&\hspace{-0.36truecm} S^{I,II}(\{u_{\a}\};\{v_i\})\no\\
     &=&\hspace{-0.36truecm} \langle 1,\ldots,1|\tilde{\T}^-_F
     (\l\hspace{-0.1truecm}-\hspace{-0.1truecm}2(M\hspace{-0.1truecm}-\hspace{-0.1truecm}1)
     \eta\hat{1},\l|\bar{u}_{N})^2_1 \ldots
     \tilde{\T}^-_F(\l\hspace{-0.1truecm}+\hspace{-0.1truecm}2M\eta\hat{1},\l|\bar{u}_1)^2_1 |2,\ldots,2\rangle,\\
 {\cal Z}^{(II)}_N(\{\bar{u}_{J}\})\hspace{-0.36truecm}&\equiv& \hspace{-0.36truecm}
     S^{II,I}(\{u_{\a}\};\{v_i\})\no\\
     &=&\hspace{-0.36truecm}\langle 2,\ldots,2|\tilde{\T}^+_F(\l,\l\hspace{-0.1truecm}+\hspace{-0.1truecm}
     2(M\hspace{-0.1truecm}-\hspace{-0.1truecm}1)\eta\hat{2}|\bar{u}_{N})^1_2\ldots
     \tilde{\T}^+_F(\l,\l\hspace{-0.1truecm}-\hspace{-0.1truecm}2M\eta\hat{2}|\bar{u}_1)^1_2|1,\ldots,1\rangle,
\eea where $N$ free parameters $\{\bar{u}_J|J=1,\ldots N\}$ are given by
\bea
\bar{u}_i=u_i\,\, {\rm for}\,\, i=1,\ldots M,\qquad {\rm and}\qquad\bar{u}_{M+i}=v_i \,\, {\rm for}\,\, i=1,\ldots M.
\eea  Note that these
functions ${\cal Z}^{(I)}_N(\{\bar{u}_{J}\})$ and ${\cal Z}^{(II)}_N(\{\bar{u}_{J}\})$ correspond to the partition functions
of the eight-vertex model with domain wall
boundary conditions and one reflecting end \cite{Tsu98} specified by the non-diagonal K-matrices
(\ref{K-matrix}) and (\ref{DK-matrix})
respectively \cite{Yan11}.

The polarization free representations
(\ref{Creation-operator-1}) and (\ref{Creation-operator-2}) of the
pseudo-particle creation/annihilation operators allowed ones \cite{Yan11} to express the above functions in terms of the
determinants representations of some $N\times N$ matrices as follows:
\bea
{\cal Z}^{(I)}_N(\{\bar{u}_{J}\})&=&\prod_{k=1}^M\frac{\s(\l_{12}+2k\eta)\s(\l_{12}-2k\eta+\eta)}
  {\s(\l_{12}+k\eta)\s(\l_{12}-k\eta+\eta)}
  \prod_{l=1}^N\prod_{i=1}^N\frac{\s(\bar{u}_i+z_i)}{\s(\bar{u}_i+z_i+\eta)}\no\\
  &&\quad \times \frac{\prod_{\a=1}^N\prod_{i=1}^N\s(\bar{u}_{\a}-z_i)\s(\bar{u}_{\a}+z_i+\eta)
   {\rm det}{\cal N}^{(I)}(\{\bar{u}_{\a}\};\{z_i\})}
  {\prod_{\a>\b}\s(\bar{u}_{\a}\hspace{-0.1truecm}-\hspace{-0.1truecm}
  \bar{u}_{\b})\s(\bar{u}_{\a}\hspace{-0.1truecm}+\hspace{-0.1truecm}\bar{u}_{\b}
  \hspace{-0.1truecm}+\hspace{-0.1truecm}\eta)\prod_{k<l}
  \s(z_k\hspace{-0.1truecm}-\hspace{-0.1truecm}z_l)\s(z_k\hspace{-0.1truecm}+\hspace{-0.1truecm}z_l)},
  \label{partition-1}\\
{\cal Z}^{(II)}_N(\{\bar{u}_{J}\})&=&\prod_{k=1}^M\frac{\s(\l_{21}+\eta-2k\eta)\s(\l_{21}-\eta+2k\eta)}
  {\s(\l_{21}-k\eta)\s(\l_{21}+k\eta-\eta)}
  \prod_{l=1}^N\prod_{i=1}^N\frac{\s(\bar{u}_i+z_i)}{\s(\bar{u}_i+z_i+\eta)}\no\\
  &&\quad \times \frac{\prod_{\a=1}^N\prod_{i=1}^N\s(\bar{u}_{\a}+z_i)\s(\bar{u}_{\a}-z_i+\eta)
   {\rm det}{\cal N}^{(II)}(\{\bar{u}_{\a}\};\{z_i\})}
  {\prod_{\a>\b}\s(\bar{u}_{\a}\hspace{-0.1truecm}-\hspace{-0.1truecm}
  \bar{u}_{\b})\s(\bar{u}_{\a}\hspace{-0.1truecm}+\hspace{-0.1truecm}\bar{u}_{\b}
  \hspace{-0.1truecm}+\hspace{-0.1truecm}\eta)\prod_{k<l}
  \s(z_l\hspace{-0.1truecm}-\hspace{-0.1truecm}z_k)\s(z_l\hspace{-0.1truecm}+\hspace{-0.1truecm}z_k)},
  \label{partition-2}
\eea where the $N\times N$ matrices  ${\cal N}^{(I)}(\{\bar{u}_{\a}\};\{z_i\})$ and
${\cal N}^{(II)}(\{\bar{u}_{\a}\};\{z_i\})$ are given by
\bea
{\cal N}^{(I)}(\{\bar{u}_{\a}\};\{z_i\})_{\a,j}&=&
  \frac{\s(\eta)\s(\l_1+\xi-z_j)}
  {\s(\bar{u}_{\a}-z_j)\s(\bar{u}_{\a}+z_j+\eta)
  \s(\l_1+\xi+\bar{u}_{\a})}\no\\
  &&\times \frac{\s(\l_2+\xi+z_j)\s(2\bar{u}_{\a})}
  {\s(\l_2+\xi+\bar{u}_{\a})
  \s(\bar{u}_{\a}-z_j+\eta)
  \s(\bar{u}_{\a}+z_j)},\\
{\cal N}^{(II)}(\{\bar{u}_{\a}\};\{z_i\})_{\a,j}&=&
  \frac{\s(\eta)\s(\l_2+\bar{\xi}-z_j)}
  {\s(\bar{u}_{\a}-z_j)\s(\bar{u}_{\a}+z_j+\eta)
  \s(\l_2+\bar{\xi}-\bar{u}_{\a}-\eta)}\no\\
  &&\times \frac{\s(\l_1+\bar{\xi}+z_j)\s(2\bar{u}_{\a}+2\eta)}
  {\s(\l_1+\bar{\xi}-\bar{u}_{\a}-\eta)
  \s(\bar{u}_{\a}-z_j+\eta)
  \s(\bar{u}_{\a}+z_j)}.
\eea  The above single determinant representations are crucial to construct
the determinant representations of the remaining scalar products $S^{I,I}$ and $S^{II,II}$ in the
next subsection.

\subsection{The scalar products $S^{I,I}$ and $S^{II,II}$}
Let us introduce two sets of  functions $\{H^{(I)}_j(u;\{z_i\},\{v_i\})|j=1,\ldots,M\}$ and
$\{H^{(II)}_j(u;\{z_i\},\{v_i\})|j=1,\ldots,M\}$
\bea
H^{(I)}_j(u;\{z_i\},\{v_i\})&=&F_1(u)\prod_{l=1}^N\frac{\s(u+z_l)}{\s(u+z_l+\eta)}
      \frac{\prod_{k\neq j}\s(u+v_k+2\eta)\s(u-v_k+\eta)}
      {\s(u-v_j)\s(u+v_j+\eta)\s(2u+\eta)}\no\\
      &&-F_2(u)\hspace{-0.1truecm}\prod_{l=1}^N\hspace{-0.1truecm}
      \frac{\s(u-z_l+\eta)}{\s(u-z_l)}
      \frac{\prod_{k\neq j}\s(u+v_k)\s(u-v_k-\eta)}
      {\s(u\hspace{-0.1truecm}-\hspace{-0.1truecm}v_j)
      \s(u\hspace{-0.1truecm}+\hspace{-0.1truecm}v_j
      \hspace{-0.1truecm}+\hspace{-0.1truecm}\eta)
      \s(2u\hspace{-0.1truecm}+\hspace{-0.1truecm}\eta)},\\
H^{(II)}_j(u;\{z_i\},\{v_i\})&=&F_3(u)\prod_{l=1}^N\frac{\s(u-z_l)}{\s(u-z_l+\eta)}
      \frac{\prod_{k\neq j}\s(v_k+u+2\eta)\s(v_k-u-\eta)}
       {\s(u+v_j+\eta)\s(u-v_j)\s(2u+\eta)}\no\\
      &&-F_4(u)\hspace{-0.1truecm}\prod_{l=1}^N\hspace{-0.1truecm}
      \frac{\s(u+z_l+\eta)}{\s(u+z_l)}
      \frac{\prod_{k\neq j}\s(v_k+u)\s(v_k-u+\eta)}
       {\s(u\hspace{-0.1truecm}+\hspace{-0.1truecm}v_j
       \hspace{-0.1truecm}+\hspace{-0.1truecm}\eta)
       \s(u\hspace{-0.1truecm}-\hspace{-0.1truecm}v_j)
       \s(2u\hspace{-0.1truecm}+\hspace{-0.1truecm}\eta)},
\eea  where the coefficients $\{F_i(u)|i=1,2,3,4\}$ are
\bea
 F_1(u)&=& \s(\l_2\hspace{-0.1truecm}+\hspace{-0.1truecm}\bar{\xi}
     \hspace{-0.1truecm}+\hspace{-0.1truecm}u
     \hspace{-0.1truecm}+\hspace{-0.1truecm}\eta)
     \s(\l_2\hspace{-0.1truecm}+\hspace{-0.1truecm}\xi
     \hspace{-0.1truecm}-\hspace{-0.1truecm}u\hspace{-0.1truecm}-\hspace{-0.1truecm}\eta)
     \s(\l_1\hspace{-0.1truecm}+\hspace{-0.1truecm}\bar{\xi}\hspace{-0.1truecm}-\hspace{-0.1truecm}
     u\hspace{-0.1truecm}-\hspace{-0.1truecm}\eta)
     \s(\l_1\hspace{-0.1truecm}+\hspace{-0.1truecm}\xi\hspace{-0.1truecm}+\hspace{-0.1truecm}u
     \hspace{-0.1truecm}+\hspace{-0.1truecm}\eta),\\
 F_2(u)&=&\s(\l_2+\bar{\xi}-u)\s(\l_2+\xi+u)
     \s(\l_1+\bar{\xi}+u)\s(\l_1+\xi-u),\\
 F_3(u)&=& \s(\l_2\hspace{-0.1truecm}+\hspace{-0.1truecm}\bar{\xi}
     \hspace{-0.1truecm}-\hspace{-0.1truecm}u\hspace{-0.1truecm}-\hspace{-0.1truecm}\eta)
     \s(\l_2\hspace{-0.1truecm}+\hspace{-0.1truecm}\xi\hspace{-0.1truecm}+\hspace{-0.1truecm}u
     \hspace{-0.1truecm}+\hspace{-0.1truecm}\eta)
     \s(\l_1\hspace{-0.1truecm}+\hspace{-0.1truecm}\bar{\xi}\hspace{-0.1truecm}+\hspace{-0.1truecm}u\hspace{-0.1truecm}
     +\hspace{-0.1truecm}\eta)
     \s(\l_1\hspace{-0.1truecm}+\hspace{-0.1truecm}\xi
     \hspace{-0.1truecm}-\hspace{-0.1truecm}u\hspace{-0.1truecm}-\hspace{-0.1truecm}\eta),\\
 F_4(u)&=&\s(\l_2+\bar{\xi}+u)\s(\l_2+\xi-u)
     \s(\l_1+\bar{\xi}-u)\s(\l_1+\xi+u).
\eea

Let us consider the scalar product $S^{I,I}(\{u_{\a}\};\{v^{(1)}_i\})$ defined by (\ref{Scalar-2}). The expression (\ref{Scalar-6})
of $S^{I,I}(\{u_{\a}\};\{v^{(1)}_i\})$ under the F-basis and the polarization free representations
(\ref{Creation-operator-1}) and (\ref{Creation-operator-2}) of the
pseudo-particle creation/annihilation operators allow us to compute the scalar product following the similar procedure as that in
\cite{Kit99} for the bulk case as follows. In front of each operators $\tilde{\T}^-_F$ in (\ref{Scalar-6}), we insert
a sum over the complete set of spin states $|j_1,\ldots,j_i\gg$, where $|j_1,\ldots,j_i\gg$ is the state with $i$ spins
being $\e_2$  in the sites $j_1,\ldots,j_i$ and $2M-i$ spins being $\e_1$ in the other sites. We are thus led to consider
some intermediate functions of the form
\bea
 G^{(i)}(u_1,\ldots,u_i|j_{i+1},\ldots,j_M;\{v^{(1)}_i\})
    \hspace{-0.38truecm}&=&\hspace{-0.38truecm}
    \ll \hspace{-0.1truecm} j_{i+1},\ldots,j_M \hspace{-0.1truecm}|
    \tilde{\T}^-_F(\l\hspace{-0.1truecm}-\hspace{-0.1truecm}2(i\hspace{-0.1truecm}-\hspace{-0.1truecm}1)
    \eta\hat{1},\l|u_i)^2_1\ldots \tilde{\T}^-_F(\l,\l|u_1)^2_1\no\\
 \hspace{-0.38truecm}&&\hspace{-0.38truecm}\times \tilde{\T}^+_F(\l,\l
    \hspace{-0.1truecm}+\hspace{-0.1truecm}2\eta\hat{1}|v^{(1)}_1)^1_2 \ldots
    \tilde{\T}^+_F(\l,\l\hspace{-0.1truecm}+\hspace{-0.1truecm}2M\eta\hat{1}|v^{(1)}_M)^1_2|1,\ldots,1\rangle,\no\\
 &&\qquad\qquad\qquad\qquad\qquad\qquad i=0,1,\ldots,M,
\eea which satisfy the following recursive relation:
\bea
  &&G^{(i)}(u_1,\ldots,u_i|j_{i+1},\ldots,j_M;\{v^{(1)}_i\})\no\\
  &&\qquad\qquad=\sum_{j\neq j_{i+1},\ldots,j_M}\ll \hspace{-0.1truecm} j_{i+1},\ldots,j_M \hspace{-0.1truecm}|
    \tilde{\T}^-_F(\l\hspace{-0.1truecm}-\hspace{-0.1truecm}2(i\hspace{-0.1truecm}-\hspace{-0.1truecm}1)
    \eta\hat{1},\l|u_i)^2_1|j,j_{i+1},\ldots,j_M\hspace{-0.1truecm}\gg\no\\
  &&\quad\quad\quad\qquad\qquad\times G^{(i-1)}(u_1,\ldots,u_{i-1}|j,j_{i+1},\ldots,j_M;\{v^{(1)}_i\}),
  \quad i=1,\ldots,M.\label{Recursive}
\eea Note that the last of these functions $\{G^{(i)}|i=0,\ldots,M\}$ is precisely the scalar product $S^{I,I}(\{u_{\a}\};\{v^{(1)}_i\})$,
namely,
\bea
 G^{(M)}(u_1,\ldots,u_M;\{v^{(1)}_i\})=S^{I,I}(\{u_{\a}\};\{v^{(1)}_i\}),
\eea whereas the first one,
\bea
 G^{(0)}(j_{1},\ldots,j_M;\{v^{(1)}_i\})=\ll \hspace{-0.1truecm} j_{1},\ldots,j_M \hspace{-0.1truecm}|
  \tilde{\T}^+_F(\l,\l\hspace{-0.1truecm}+\hspace{-0.1truecm}2\eta\hat{1}|v^{(1)}_1)^1_2
  \ldots\tilde{\T}^+_F(\l,\l\hspace{-0.1truecm}+\hspace{-0.1truecm}2M\eta\hat{1}|v^{(1)}_M)^1_2
  |1,\ldots,1\rangle,\no
\eea  is closely related to the partition function computed in \cite{Yan11}. Solving the recursive relations
(\ref{Recursive}), we find that
if the parameters $\{v^{(1)}_k\}$ satisfy the first set of  Bethe
ansatz equations (\ref{BA-D-1}) the scalar product
$S^{I,I}(\{u_{\a}\}; \{v^{(1)}_i\})$ has the following determinant representation
\bea
S^{I,I}(\{u_{\a}\};\{v^{(1)}_i\})
  \hspace{-0.38truecm}&=&\hspace{-0.4truecm}\prod_{k=1}^M\hspace{-0.1truecm}\lt\{\hspace{-0.1truecm}
  \frac{\s(\l_{12}\hspace{-0.1truecm}+\hspace{-0.1truecm}2\eta\hspace{-0.1truecm}-\hspace{-0.1truecm}2k\eta)
  \s(\l_{12}\hspace{-0.1truecm}-\hspace{-0.1truecm}\eta\hspace{-0.1truecm}+\hspace{-0.1truecm}2k\eta)}
  {\s(\l_{12}\hspace{-0.1truecm}-\hspace{-0.1truecm}
  (k\hspace{-0.1truecm}-\hspace{-0.1truecm}1)\eta)
  \s(\l_{12}\hspace{-0.1truecm}+\hspace{-0.1truecm}k\eta)}\hspace{-0.1truecm}
  \prod_{l=1}^{N}\hspace{-0.1truecm}\frac{\s(u_k-z_l)\s(v^{(1)}_k-z_l)}
  {\s(u_k\hspace{-0.1truecm}-\hspace{-0.1truecm}z_l\hspace{-0.1truecm}+\hspace{-0.1truecm}\eta)
  \s(v^{(1)}_k\hspace{-0.1truecm}-\hspace{-0.1truecm}z_l\hspace{-0.1truecm}+\hspace{-0.1truecm}\eta)}
  \hspace{-0.1truecm}\rt\}\no\\
\hspace{-0.38truecm}&&\hspace{-0.28truecm}\times\hspace{-0.1truecm}
  \frac{{\rm det}\bar{{\cal N}}^{(I)}(\{u_{\a}\};\{v^{(1)}_i\})}
  {\prod_{\a<\b}\s(u_{\a}
  \hspace{-0.1truecm}-\hspace{-0.1truecm}u_{\b})
  \s(u_{\a}\hspace{-0.1truecm}+\hspace{-0.1truecm}u_{\b}
  \hspace{-0.1truecm}+\hspace{-0.1truecm}\eta)
  \hspace{-0.1truecm}\prod_{k>l}\s(v^{(1)}_k
  \hspace{-0.1truecm}-\hspace{-0.1truecm}v^{(1)}_l)\s(v^{(1)}_k
  \hspace{-0.1truecm}+\hspace{-0.1truecm}v^{(1)}_l
  \hspace{-0.1truecm}+\hspace{-0.1truecm}\eta)},\no\\
  \label{Determinant-1}
\eea where the $M\times M$ matrix $\bar{{\cal N}}^{(I)}(\{u_{\a}\};\{v^{(1)}_i\})$ is given by
\bea
 \bar{{\cal N}}^{(I)}(\{u_{\a}\};\{v^{(1)}_i\})_{\a,j}\hspace{-0.1truecm}=\hspace{-0.1truecm}\frac
 {\s(\eta)\s(2u_{\a})\s(2v^{(1)}_j+2\eta)H^{(I)}_j(u_{\a};\{z_i\},\{v^{(1)}_i\})}
 {\s(\l_1\hspace{-0.1truecm}+\hspace{-0.1truecm}\xi
 \hspace{-0.1truecm}+\hspace{-0.1truecm}u_{\a})
 \s(\l_2\hspace{-0.1truecm}+\hspace{-0.1truecm}\xi\hspace{-0.1truecm}+\hspace{-0.1truecm}u_{\a})
 \s(\l_2\hspace{-0.1truecm}+\hspace{-0.1truecm}\bar{\xi}
 \hspace{-0.1truecm}-\hspace{-0.1truecm}v^{(1)}_j
 \hspace{-0.1truecm}-\hspace{-0.1truecm}\eta)\s(\l_1\hspace{-0.1truecm}+
 \hspace{-0.1truecm}\bar{\xi}\hspace{-0.1truecm}-\hspace{-0.1truecm}v^{(1)}_{j}
 \hspace{-0.1truecm}-\hspace{-0.1truecm}\eta)}.\no\\
\eea Using the similar method as above, we have that the scalar product
$S^{II,II}(\{u_{\a}\}; \{v^{(2)}_i\})$ has the following determinant representation provided that
the parameters  $\{v^{(2)}_k\}$ satisfy the second set of  Bethe
ansatz equations (\ref{BA-D-2})
\bea
S^{II,II}(\{u_{\a}\};\{v^{(2)}_i\})
  \hspace{-0.38truecm}&=&\hspace{-0.4truecm}\prod_{k=1}^M\hspace{-0.1truecm}\lt\{\hspace{-0.1truecm}
  \frac{\s(\l_{12}\hspace{-0.1truecm}+\hspace{-0.1truecm}2k\eta)
  \s(\l_{21}\hspace{-0.1truecm}-\hspace{-0.1truecm}\eta
  \hspace{-0.1truecm}+\hspace{-0.1truecm}2k\eta)}
  {\s(\l_{12}\hspace{-0.1truecm}+\hspace{-0.1truecm}k\eta)
  \s(\l_{21}\hspace{-0.1truecm}+\hspace{-0.1truecm}(k\hspace{-0.1truecm}-\hspace{-0.1truecm}1)\eta)}\hspace{-0.1truecm}
  \prod_{l=1}^{N}\hspace{-0.1truecm}\frac{\s(u_k+z_l)\s(v^{(2)}_k+z_l)}
  {\s(u_k\hspace{-0.1truecm}+\hspace{-0.1truecm}z_l\hspace{-0.1truecm}+\hspace{-0.1truecm}\eta)
  \s(v^{(2)}_k\hspace{-0.1truecm}+\hspace{-0.1truecm}z_l\hspace{-0.1truecm}+\hspace{-0.1truecm}\eta)}
  \hspace{-0.1truecm}\rt\}\no\\
\hspace{-0.38truecm}&&\hspace{-0.28truecm}\times\hspace{-0.1truecm}
  \frac{{\rm det}\bar{{\cal N}}^{(II)}(\{u_{\a}\};\{v^{(2)}_i\})}
  {\prod_{\a<\b}\s(u_{\a}
  \hspace{-0.1truecm}-\hspace{-0.1truecm}u_{\b})
  \s(u_{\a}\hspace{-0.1truecm}+\hspace{-0.1truecm}u_{\b}
  \hspace{-0.1truecm}+\hspace{-0.1truecm}\eta)
  \hspace{-0.1truecm}\prod_{k>l}\s(v^{(2)}_k
  \hspace{-0.1truecm}-\hspace{-0.1truecm}v^{(2)}_l)\s(v^{(2)}_k
  \hspace{-0.1truecm}+\hspace{-0.1truecm}v^{(2)}_l
  \hspace{-0.1truecm}+\hspace{-0.1truecm}\eta)},\no\\
  \label{Determinant-2}
\eea where the $M\times M$ matrix $\bar{{\cal N}}^{(II)}(\{u_{\a}\};\{v^{(2)}_i\})$ is given by
\bea
 \bar{{\cal N}}^{(II)}(\{u_{\a}\};\{v^{(2)}_i\})_{\a,j}\hspace{-0.1truecm}=\hspace{-0.1truecm}\frac
 {\s(\eta)\s(2u_{\a}+2\eta)\s(2v^{(2)}_j)H^{(II)}_j(u_{\a};\{z_i\},\{v^{(2)}_i\})}
 {\s(\l_2\hspace{-0.1truecm}+\hspace{-0.1truecm}\bar{\xi}
 \hspace{-0.1truecm}-\hspace{-0.1truecm}u_{\a}\hspace{-0.1truecm}-\hspace{-0.1truecm}\eta)
 \s(\l_1\hspace{-0.1truecm}+\hspace{-0.1truecm}\bar{\xi}\hspace{-0.1truecm}-\hspace{-0.1truecm}u_{\a}
 \hspace{-0.1truecm}-\hspace{-0.1truecm}\eta)
 \s(\l_2\hspace{-0.1truecm}+\hspace{-0.1truecm}\xi
 \hspace{-0.1truecm}+\hspace{-0.1truecm}v^{(2)}_j)\s(\l_1\hspace{-0.1truecm}+
 \hspace{-0.1truecm}\xi\hspace{-0.1truecm}+\hspace{-0.1truecm}v^{(2)}_{j})}.\no\\
\eea

Now we are in position to compute the norms of the Bethe states which can be obtained
by taking the limit $u_{\a}\rightarrow v^{(i)}_{\a},$ $\a=1,\ldots M$. The norm of the first set of
Bethe state (\ref{Bethe-state-1}) is
\bea
 \mathbb{N}^{I,I}(\{v^{(1)}_{\a}\})&=&\lim_{u_{\a}\rightarrow v^{(1)}_{\a}}S^{I,I}(\{u_{\a}\}; \{v^{(1)}_i\})\no\\[6pt]
   &=&\prod_{k=1}^M\hspace{-0.1truecm}\lt\{\hspace{-0.1truecm}
  \frac{\s(\l_{12}\hspace{-0.1truecm}+\hspace{-0.1truecm}2\eta\hspace{-0.1truecm}-\hspace{-0.1truecm}2k\eta)
  \s(\l_{12}\hspace{-0.1truecm}-\hspace{-0.1truecm}\eta\hspace{-0.1truecm}+\hspace{-0.1truecm}2k\eta)}
  {\s(\l_{12}\hspace{-0.1truecm}-\hspace{-0.1truecm}
  (k\hspace{-0.1truecm}-\hspace{-0.1truecm}1)\eta)
  \s(\l_{12}\hspace{-0.1truecm}+\hspace{-0.1truecm}k\eta)}\hspace{-0.1truecm}
  \prod_{l=1}^{N}\hspace{-0.1truecm}\frac{\s^2(v^{(1)}_k-z_l)}
  {\s^2(v^{(1)}_k\hspace{-0.1truecm}-\hspace{-0.1truecm}z_l\hspace{-0.1truecm}+\hspace{-0.1truecm}\eta)}
  \hspace{-0.1truecm}\rt\}\no\\[6pt]
  &&\times \prod_{\a\ne \b}\frac{\s(v^{(1)}_{\a}+v^{(1)}_{\b})\s(v^{(1)}_{\a}-v^{(1)}_{\b}-\eta)}
  {\s(v^{(1)}_{\a}-v^{(1)}_{\b})\s(v^{(1)}_{\a}+v^{(1)}_{\b}+\eta)} \, {\rm det}\Phi^{(I)}(\{v^{(1)}_{\a}\}),
  \label{Norm-1}
\eea where the matrix elements of $M\times M$ matrix $\Phi^{(I)}(\{v_{\a}\})$ are given by
\bea
  \Phi^{(I)}_{\a,j}(\{v_{\a}\})&=& \frac{\s(\eta)
     \s(\l_2\hspace{-0.1truecm}+\hspace{-0.1truecm}\bar{\xi}\hspace{-0.1truecm}-\hspace{-0.1truecm}v_{\a})
     \s(\l_1\hspace{-0.1truecm}+\hspace{-0.1truecm}\bar{\xi}\hspace{-0.1truecm}+\hspace{-0.1truecm}v_{\a})
     \s(\l_1\hspace{-0.1truecm}+\hspace{-0.1truecm}\xi-\hspace{-0.1truecm}v_{\a})}
     {\s(\l_1\hspace{-0.1truecm}+\hspace{-0.1truecm}\xi\hspace{-0.1truecm}+\hspace{-0.1truecm}v_{\a})
     \s(\l_2\hspace{-0.1truecm}+\hspace{-0.1truecm}\bar{\xi}\hspace{-0.1truecm}-\hspace{-0.1truecm}v_j\hspace{-0.1truecm}-\hspace{-0.1truecm}\eta)
     \s(\l_1\hspace{-0.1truecm}+\hspace{-0.1truecm}\bar{\xi}\hspace{-0.1truecm}-\hspace{-0.1truecm}v_j\hspace{-0.1truecm}-\hspace{-0.1truecm}\eta)}
     \no\\[6pt]
  &&\times \frac{\s(2v_{\a})\s(2v_j\hspace{-0.1truecm}+\hspace{-0.1truecm}2\eta)}
     {\s(2v_{\a}\hspace{-0.1truecm}+\hspace{-0.1truecm}\eta)\s(2v_j\hspace{-0.1truecm}+\hspace{-0.1truecm}\eta)}
     \prod_{l=1}^N\frac{\s(v_{\a}-z_l+\eta)}{\s(v_{\a}-z_l)}\no\\[6pt]
  &&\times\frac{\partial}{\partial v_{\a}}\,\ln\lt\{
     \frac{\s(\l_2\hspace{-0.1truecm}+\hspace{-0.1truecm}\bar{\xi}\hspace{-0.1truecm}+\hspace{-0.1truecm}v_j\hspace{-0.1truecm}+\hspace{-0.1truecm}\eta)
     \s(\l_2\hspace{-0.1truecm}+\hspace{-0.1truecm}\xi\hspace{-0.1truecm}-\hspace{-0.1truecm}v_j
     \hspace{-0.1truecm}-\hspace{-0.1truecm}\eta)
     \s(\l_1\hspace{-0.1truecm}+\hspace{-0.1truecm}\bar{\xi}\hspace{-0.1truecm}-\hspace{-0.1truecm}v_j
     \hspace{-0.1truecm}-\hspace{-0.1truecm}\eta)
     \s(\l_1\hspace{-0.1truecm}+\hspace{-0.1truecm}\xi+\hspace{-0.1truecm}v_j\hspace{-0.1truecm}+\hspace{-0.1truecm}\eta)}
     {\s(\l_2\hspace{-0.1truecm}+\hspace{-0.1truecm}\bar{\xi}\hspace{-0.1truecm}-\hspace{-0.1truecm}v_j)
     \s(\l_2\hspace{-0.1truecm}+\hspace{-0.1truecm}\xi\hspace{-0.1truecm}+\hspace{-0.1truecm}v_j)
     \s(\l_1\hspace{-0.1truecm}+\hspace{-0.1truecm}\bar{\xi}\hspace{-0.1truecm}+\hspace{-0.1truecm}v_j)
     \s(\l_1\hspace{-0.1truecm}+\hspace{-0.1truecm}\xi\hspace{-0.1truecm}-\hspace{-0.1truecm}v_j)}\rt.\no\\[6pt]
  &&\quad\quad\times \prod_{l=1}^N\lt.
     \frac{\s(v_j\hspace{-0.1truecm}+\hspace{-0.1truecm}z_l)\s(v_j\hspace{-0.1truecm}-\hspace{-0.1truecm}z_l)}
     {\s(v_j\hspace{-0.1truecm}+\hspace{-0.1truecm}z_l\hspace{-0.1truecm}+\hspace{-0.1truecm}\eta)
     \s(v_j\hspace{-0.1truecm}-\hspace{-0.1truecm}z_l\hspace{-0.1truecm}+\hspace{-0.1truecm}\eta)}
     \prod_{k\ne j}
     \frac{\s(v_j\hspace{-0.1truecm}+\hspace{-0.1truecm}v_k\hspace{-0.1truecm}+\hspace{-0.1truecm}2\eta)
     \s(v_j\hspace{-0.1truecm}-\hspace{-0.1truecm}v_k\hspace{-0.1truecm}+\hspace{-0.1truecm}\eta)}
     {\s(v_j+v_k)\s(v_j\hspace{-0.1truecm}-\hspace{-0.1truecm}v_k-\eta)}\rt\},
     \label{Norm-2}
\eea the norm of the second set of
Bethe state (\ref{Bethe-state-2}) is given by
\bea
 \mathbb{N}^{II,II}(\{v^{(2)}_{\a}\})&=&\lim_{u_{\a}\rightarrow v^{(2)}_{\a}}S^{II,II}(\{u_{\a}\}; \{v^{(2)}_i\})\no\\[6pt]
   &=&\prod_{k=1}^M\hspace{-0.1truecm}\lt\{\hspace{-0.1truecm}
  \frac{\s(\l_{12}\hspace{-0.1truecm}+\hspace{-0.1truecm}2k\eta)
  \s(\l_{21}\hspace{-0.1truecm}-\hspace{-0.1truecm}\eta\hspace{-0.1truecm}+\hspace{-0.1truecm}2k\eta)}
  {\s(\l_{12}\hspace{-0.1truecm}+\hspace{-0.1truecm}k\eta)
  \s(\l_{21}\hspace{-0.1truecm}+\hspace{-0.1truecm}(k\hspace{-0.1truecm}-\hspace{-0.1truecm}1)\eta)}\hspace{-0.1truecm}
  \prod_{l=1}^{N}\hspace{-0.1truecm}\frac{\s^2(v^{(2)}_k+z_l)}
  {\s^2(v^{(2)}_k\hspace{-0.1truecm}+\hspace{-0.1truecm}z_l\hspace{-0.1truecm}+\hspace{-0.1truecm}\eta)}
  \hspace{-0.1truecm}\rt\}\no\\[6pt]
  &&\times \prod_{\a\ne \b}\frac{\s(v^{(2)}_{\a}\hspace{-0.1truecm}+\hspace{-0.1truecm}v^{(2)}_{\b}
  \hspace{-0.1truecm}+\hspace{-0.1truecm}2\eta)\s(v^{(2)}_{\a}\hspace{-0.1truecm}-\hspace{-0.1truecm}v^{(2)}_{\b}
  \hspace{-0.1truecm}-\hspace{-0.1truecm}\eta)}
  {\s(v^{(2)}_{\a}\hspace{-0.1truecm}-\hspace{-0.1truecm}v^{(2)}_{\b})\s(v^{(2)}_{\a}
  \hspace{-0.1truecm}+\hspace{-0.1truecm}v^{(2)}_{\b}\hspace{-0.1truecm}+\hspace{-0.1truecm}\eta)} \, {\rm det}\Phi^{(II)}(\{v^{(2)}_{\a}\}),
  \label{Norm-3}
\eea where the matrix elements of $M\times M$ matrix $\Phi^{(II)}(\{v_{\a}\})$ are given by
\bea
  \Phi^{(II)}_{\a,j}(\{v_{\a}\})&=& \frac{\s(\eta)
     \s(\l_2\hspace{-0.1truecm}+\hspace{-0.1truecm}\xi\hspace{-0.1truecm}+\hspace{-0.1truecm}v_{\a}\hspace{-0.1truecm}+\hspace{-0.1truecm}\eta)
     \s(\l_1\hspace{-0.1truecm}+\hspace{-0.1truecm}\bar{\xi}\hspace{-0.1truecm}+\hspace{-0.1truecm}v_{\a}\hspace{-0.1truecm}+\hspace{-0.1truecm}\eta)
     \s(\l_1\hspace{-0.1truecm}+\hspace{-0.1truecm}\xi-\hspace{-0.1truecm}v_{\a}\hspace{-0.1truecm}-\hspace{-0.1truecm}\eta)}
     {\s(\l_1\hspace{-0.1truecm}+\hspace{-0.1truecm}\bar{\xi}\hspace{-0.1truecm}-\hspace{-0.1truecm}v_{\a}\hspace{-0.1truecm}-\hspace{-0.1truecm}\eta)
     \s(\l_2\hspace{-0.1truecm}+\hspace{-0.1truecm}\xi\hspace{-0.1truecm}+\hspace{-0.1truecm}v_j)
     \s(\l_1\hspace{-0.1truecm}+\hspace{-0.1truecm}\xi\hspace{-0.1truecm}+\hspace{-0.1truecm}v_j)}
     \no\\[6pt]
  &&\times \frac{\s(2v_{\a}\hspace{-0.1truecm}+\hspace{-0.1truecm}2\eta)\s(2v_j)}
     {\s(2v_{\a}\hspace{-0.1truecm}+\hspace{-0.1truecm}\eta)\s(2v_j\hspace{-0.1truecm}+\hspace{-0.1truecm}\eta)}
     \prod_{l=1}^N\frac{\s(v_{\a}-z_l)}{\s(v_{\a}-z_l+\eta)}\no\\[6pt]
  &&\times\frac{\partial}{\partial v_{\a}}\,\ln\lt\{
     \frac{\s(\l_2\hspace{-0.1truecm}+\hspace{-0.1truecm}\bar{\xi}\hspace{-0.1truecm}-\hspace{-0.1truecm}v_j\hspace{-0.1truecm}-\hspace{-0.1truecm}\eta)
     \s(\l_2\hspace{-0.1truecm}+\hspace{-0.1truecm}\xi\hspace{-0.1truecm}+\hspace{-0.1truecm}v_j
     \hspace{-0.1truecm}+\hspace{-0.1truecm}\eta)
     \s(\l_1\hspace{-0.1truecm}+\hspace{-0.1truecm}\bar{\xi}\hspace{-0.1truecm}+\hspace{-0.1truecm}v_j
     \hspace{-0.1truecm}+\hspace{-0.1truecm}\eta)
     \s(\l_1\hspace{-0.1truecm}+\hspace{-0.1truecm}\xi-\hspace{-0.1truecm}v_j\hspace{-0.1truecm}-\hspace{-0.1truecm}\eta)}
     {\s(\l_2\hspace{-0.1truecm}+\hspace{-0.1truecm}\bar{\xi}\hspace{-0.1truecm}+\hspace{-0.1truecm}v_j)
     \s(\l_2\hspace{-0.1truecm}+\hspace{-0.1truecm}\xi\hspace{-0.1truecm}-\hspace{-0.1truecm}v_j)
     \s(\l_1\hspace{-0.1truecm}+\hspace{-0.1truecm}\bar{\xi}\hspace{-0.1truecm}-\hspace{-0.1truecm}v_j)
     \s(\l_1\hspace{-0.1truecm}+\hspace{-0.1truecm}\xi\hspace{-0.1truecm}+\hspace{-0.1truecm}v_j)}\rt.\no\\[6pt]
  &&\quad\quad\times \prod_{l=1}^N\lt.
     \frac{\s(v_j\hspace{-0.1truecm}+\hspace{-0.1truecm}z_l)\s(v_j\hspace{-0.1truecm}-\hspace{-0.1truecm}z_l)}
     {\s(v_j\hspace{-0.1truecm}+\hspace{-0.1truecm}z_l\hspace{-0.1truecm}+\hspace{-0.1truecm}\eta)
     \s(v_j\hspace{-0.1truecm}-\hspace{-0.1truecm}z_l\hspace{-0.1truecm}+\hspace{-0.1truecm}\eta)}
     \prod_{k\ne j}
     \frac{\s(v_j\hspace{-0.1truecm}+\hspace{-0.1truecm}v_k\hspace{-0.1truecm}+\hspace{-0.1truecm}2\eta)
     \s(v_j\hspace{-0.1truecm}-\hspace{-0.1truecm}v_k\hspace{-0.1truecm}+\hspace{-0.1truecm}\eta)}
     {\s(v_j+v_k)\s(v_j\hspace{-0.1truecm}-\hspace{-0.1truecm}v_k-\eta)}\rt\}.
     \label{Norm-4}
\eea Moreover, one may check that if the parameters $\{u_{\a}\}$ satisfy the Bethe ansatz equations (i.e. on shell) but different
from $\{v^{(i)}_{\a}\}$ the corresponding scalar products $S^{I,I}(\{u_{\a}\}; \{v^{(1)}_i\})$ or $S^{II,II}(\{u_{\a}\}; \{v^{(2)}_i\})$
vanishes, namely, the corresponding Bethe states are orthogonal.


\section{ Conclusions}
\label{C} \setcounter{equation}{0}

We have studied scalar products between an
on-shell Bethe state and  a general state (or an off-shell Bethe state) of
the open XYZ chain with non-diagonal boundary
terms, where the non-diagonal K-matrices $K^{\pm}(u)$ are given by
(\ref{K-matrix-2-1}) and (\ref{K-matrix-6}).  In our calculation
the factorizing F-matrix (\ref{F-matrix}) of the
eight-vertex SOS model, which leads to the polarization free representations
(\ref{Creation-operator-1}) and (\ref{Creation-operator-2}) of the associated
pseudo-particle creation/annihilation  operators, has played an important role. It is found that  the scalar
products can be expressed in terms of the determinants (\ref{partition-1}), (\ref{partition-2}),
(\ref{Determinant-1}) and (\ref{Determinant-2}). By taking the on shell limit, we
obtain the determinant representations (or Gaudin formula) (\ref{Norm-1})-(\ref{Norm-2}) and (\ref{Norm-3})-(\ref{Norm-4})
of the norms of the Bethe states. However it should be emphasized that, in contrast with those of the
open XXZ chain with diagonal boundary terms, the dual states (\ref{Dual-1})
and (\ref{Dual-2}) are generally no longer the eigenstates  of the open chain with non-diagonal
boundary terms even if the parameters satisfy the associated Bethe ansatz equations.

Now let us consider the various trigonometric limits of our results. The definition of the elliptic
function (\ref{Function-a-b}) implies the following asymptotic behaviors
\bea
 &&\lim_{\tau\rightarrow+i\infty}\s(u)=-2e^{\frac{i\pi\tau}{4}}\sin\pi u+\cdots,\no\\
 &&\lim_{\tau\rightarrow+i\infty}\theta\lt[\begin{array}{c}0\\\frac{1}{2}\end{array}\rt](u,\tau)=1+\cdots.\no
\eea Let consider the first trigonometric limit, i.e. redefining the parameters as follows
\bea
 &&u\longrightarrow \frac{u}{\pi},\qquad\qquad \l_i \longrightarrow \frac{\l_i-\frac{1}{4}+\frac{3\tau}{4}}{\pi},\,i=1,2,\label{Limit-1}\\
 &&\xi\longrightarrow \frac{\xi+\frac{1}{4}-\frac{3\tau}{4}}{\pi},\quad \bar{\xi}\longrightarrow \frac{\bar{\xi}+\frac{1}{4}-\frac{3\tau}{4}}{\pi}.
    \label{Limit-2}
\eea Then taking the limit $\tau\longrightarrow +i\infty$, the resulting K-matrices $K^{\pm}(u)$ given by (\ref{K-matrix-2-1}) and (\ref{K-matrix-6})
become the non-diagonal trigonometric K-matrices considered in \cite{Yan11-1}, and our results (\ref{partition-1}), (\ref{partition-2}),
(\ref{Determinant-1}), (\ref{Determinant-2}),(\ref{Norm-1})-(\ref{Norm-2}) and (\ref{Norm-3})-(\ref{Norm-4}) recover those of \cite{Yan11-1}
for the open XXZ chain with
non-diagonal boundary terms. On the other hand, if we redefine the parameters as follows (c.f. (\ref{Limit-1})-(\ref{Limit-2}))
\bea
 &&\xi\longrightarrow \frac{\xi-\l_2+\frac{i}{2}\ln2}{\pi},\quad \bar{\xi}\longrightarrow \frac{\bar{\xi}-\l_2+\frac{i}{2}\ln2}{\pi},
    \label{Limit-3}\\
 &&u\longrightarrow \frac{u}{\pi},\label{Limit-4}
\eea and then take the limit: $\tau\longrightarrow +i\infty$ and $\l_1\longrightarrow +i\infty$, the resulting K-matrices $K^{\pm}(u)$ given by (\ref{K-matrix-2-1}) and (\ref{K-matrix-6}) become the diagonal trigonometric K-matrices and our result recovers that of \cite{Wan02,Kit07} for the open XXZ chain with diagonal boundary terms.

\section*{Acknowledgements}
The financial supports from  the National Natural Science
Foundation of China (Grant Nos. 11075126 and 11031005), Australian Research Council
are gratefully acknowledged.


\end{document}